\documentclass{article}

\usepackage{arxiv}

\usepackage[utf8]{inputenc} 
\usepackage[T1]{fontenc}    
\usepackage{hyperref}       
\usepackage{url}            
\usepackage{booktabs}       
\usepackage{amsfonts}       
\usepackage{nicefrac}       
\usepackage{microtype}      
\usepackage{lipsum}
\usepackage{graphicx}
\graphicspath{ {./images/} }

\usepackage{amsmath}
\usepackage{color}
\usepackage{bm}
\usepackage{url}
\usepackage{graphicx}
\usepackage{subfigure}
\usepackage{booktabs}
\usepackage{multirow}
\usepackage[ruled,linesnumbered]{algorithm2e}
\usepackage{amsfonts}

\title{Uncertainty-aware Frequency-domain Acoustic Full Waveform Inversion Using Gaussian Random Fields and  Ensemble Kalman Inversion}

\author{
 Yunduo Li \\
   School of Information and \\ Communications Engineering \\
  Xi’an Jiaotong University\\
  Xi’an, Shaanxi 710049, China \\
  \texttt{yunduo\_li@stu.xjtu.edu.cn} \\
   \And
 Yijie Zhang \\
   School of Information and \\ Communications Engineering \\
  Xi’an Jiaotong University\\
  Xi’an, Shaanxi 710049, China \\
  \texttt{zhangyijie2016@mail.xjtu.edu.cn} \\
  \And
 Xueyu Zhu \\
  Department of Mathematics\\
  The University of Iowa\\
  Iowa City, IA 52246, USA \\
  \texttt{xueyu-zhu@uiowa.edu} \\
  \And
 Jinghuai Gao\\
   School of Information and \\ Communications Engineering \\
  Xi’an Jiaotong University\\
  Xi’an, Shaanxi 710049, China \\
  \texttt{jhgao@mail.xjtu.edu.cn} \\
}

\begin{document}
\maketitle
\begin{abstract}
In recent years, uncertainty-aware full waveform inversion (FWI) has received increasing attention, with a growing emphasis on producing informative uncertainty estimates alongside inversion results. Bayesian inference methods—particularly Monte Carlo-based approaches—have been widely employed to quantify uncertainty. However, these techniques often require extensive posterior sampling, resulting in high computational costs.
To address this challenge and enable efficient uncertainty quantification in FWI, we introduce an uncertainty-aware FWI framework—EKI-GRFs-FWI—that integrates Gaussian random fields (GRFs) with the ensemble Kalman inversion (EKI) algorithm. This approach jointly infers subsurface velocity fields and provides reliable uncertainty estimates in a computationally efficient manner.
The EKI algorithm leverages a derivative-free update mechanism and employs effective stopping criteria to ensure rapid convergence, making it suitable for large-scale inverse problems. Meanwhile, GRFs incorporate prior knowledge of spatial smoothness and correlation length scales, enabling the generation of physically plausible initial ensembles for EKI.
Numerical results demonstrate that EKI-GRFs-FWI yields reasonably accurate velocity reconstructions while delivering informative  uncertainty estimates. 
\end{abstract}


\section{Introduction}
Full waveform inversion (FWI) is a powerful technique in geophysics that utilizes seismic wavefield data (waveform, phase and amplitude) to inverse subsurface structures. Due to its superior resolution, FWI has become a widely used method for characterizing the underground structural details.
Conventional FWI primarily relies on optimization methods, such as the Gauss-Newton \cite{https://doi.org/10.1046/j.1365-246X.1998.00498.x, 10.1093/gji/ggad299, 10.1190/geo2022-0673.1}, conjugate gradient \cite{10669381}, steepest-descent \cite{10.1190/1.3238367}, limited-memory Broyden-Fletcher-Goldfarb-Shanno (L-BFGS)  \cite{10681159, 10887266} methods.
These methods iteratively minimize the discrepancy between observed and simulated data to update the unknown physical parameters. 
Despite their ability to provide relatively accurate inverse results in certain cases, these methods face several significant challenges. For instance, optimization methods are highly sensitive to the initial model. If the initial model is far from the true model, the optimization process may converge to local minima. Furthermore, these methods typically provide an optimal solution without providing  uncertainty estimates about the inverse solution.

To address the limitations of optimization methods in uncertainty quantification, Bayesian inference method has been integrated into FWI, which provides a robust framework for incorporating prior knowledge or assumptions about the subsurface structure into the inference process \cite{10.1093/gji/ggad057, 10.1093/gji/ggae129}.
Based on Bayes' theorem, the initially assumed prior distribution is systematically updated to a posterior distribution, which offers a comprehensive estimation of the solutions to the inverse problem associated with the uncertainties. Markov Chain Monte Carlo (MCMC) is a widely used Bayesian inference approach in FWI \cite{10.1093/gji/ggab287,10.1190/geo2019-0585.1}. However, it is computationally intensive, especially in high-dimensional parameter spaces where a large number of samples are required to converge to the stationary distribution \cite{Belhadj_2018}. To alleviate the computational burden, Hamiltonian Monte Carlo (HMC) \cite{10.1093/gji/ggad403,10.1093/gji/ggae112} is proposed, which is a variant of MCMC that utilizes Hamiltonian dynamics to navigate the parameter space. By leveraging gradient information, HMC is more efficient to sample the posterior distribution \cite{10.1093/gji/ggy496}. However, the requirement on accurate gradient information by HMC leads to significant computational demands and analytical complexity, especially for large-scale or high-dimensional problems with extensive observed data.

To enhance computational efficiency and provide reasonable uncertainty estimations about the inferred results, the ensemble Kalman inversion (EKI) algorithm is introduced to solve inverse problems. EKI is a useful Bayesian inference method, which stems from the ensemble Kalman filter (EnKF) proposed by Evensen \cite{https://doi.org/10.1029/94JC00572, 2003The}.
EKI possesses several advantages.
The derivative-free nature of EKI eliminates the need for explicit gradient calculations, simplifying the computational process and rendering it suitable for non-differentiable models. Specifically, EKI uses the sample statistics of ensemble members to implicitly approximate Hessian and gradient information, where derivatives are approximated through the differences within the ensemble \cite{Zhang_Xiao_Luo_He_2022}. EKI method significantly reduces computational complexity and storage requirements, and it is highly efficient, especially for high-dimensional problems. Additionally, EKI is highly parallelizable due to the independent computations performed on multiple ensemble members. 
These characteristics make EKI perform effectively in a variety of geophysical fields, demonstrating its efficacy in applications such as electrical resistivity imaging \cite{10.1093/gji/ggab013}, traveltime tomography \cite{10.1093/gji/ggae329, doi:10.1190/geo2023-0493.1, 10.1093/gji/ggz472}. Furthermore, an ensemble-based Kalman filter method has been applied to full waveform inversion in time-domain \cite{WANG2019321}. In addition, 
ensemble transform Kalman filter algorithm is applied to frequency-domain full waveform inversion; however, it relies on adjoint-based computations to obtain the gradient \cite{10.1093/gji/ggz384}, and therefore is not a gradient-free approach.
Notably, recent advancements highlight its effectiveness in uncertainty quantification for high-dimensional problems, especially when integrated with neural networks \cite{pensoneault2023efficient, PENSONEAULT2025113670}, further emphasizing its broad utility.

Constructing an appropriately designed initial ensemble of parameters is beneficial for EKI algorithm, as it ensures adequate representation of the parameter space. The initial ensemble of parameters determines the starting point and exploration scope of parameter space  in EKI algorithm. Generally, the initial ensemble of EKI is randomly sampled from a Gaussian distribution. However, such sampling strategy often ignores the spatial correlation between parameters, which may lead to insufficient exploration of parameter space by the EKI algorithm.  To address this issue, Gaussian random fields (GRFs) is an effective method to generate the initial ensemble of parameters for EKI \cite{Iglesias_2016, Chada_2018, Iglesias_2022, 10.1093/gji/ggaf060}.
GRFs offer a flexible framework for modeling spatially correlated parameters for inverse problems \cite{Lassi}. 
By incorporating prior knowledge of the length-scale and smoothness, GRFs is able to generate an initial ensemble that captures the spatial correlation characteristics of the parameters \cite{4847c1a8-518e-3c37-8c1a-780bcc44d236}. The combination of GRFs with EKI has been successfully applied in various geophysical domains \cite{10.1093/gji/ggab013,10.1093/gji/ggae329,10.1093/gji/ggz472}, effectively demonstrating its versatility and applicability. 

In this study, we propose an efficient uncertainty-aware frequency-domain acoustic FWI method that combines the EKI algorithm with GRFs. 
The proposed method can efficiently infer relatively accurate velocity fields and provide informative uncertainty about the inverse results. 
The contributions of this work can be summarized as follows:

\begin{itemize}
    \item We utilize EKI algorithm to efficiently infer velocity fields and provide informative uncertainty for the inverse results.

    \item We adopt GRFs to generate the initial ensemble for EKI, and investigate the impact of
    GRFs with different initial length-scale on the inferred velocity fields.

    \item By incorporating prior knowledge of the length-scale in GRFs, the proposed method reduces the dependence on the initial velocity model.

    \item We propose a novel mini-batch strategy for EKI, which reduces computational costs and excessive memory usage in computing sample covariance matrices.

\end{itemize}

The rest of the paper is organized as follows. First, we introduce the frequency-domain acoustic wave equation. Second, we briefly review the EKI algorithm and Gaussian Random Fields (GRFs), then present the combination of EKI and GRFs to solve FWI problems (EKI-GRFs-FWI), as well as the mini-batch strategy. Finally, we compare the performance of EKI-GRFs-FWI with MCMC method on several benchmark velocity models.

\section{Problem Setup}
\label{sec:problem}
In this study, we use two-dimensional frequency-domain acoustic wave equation \cite{10.1093/gji/ggac399, 10017252} to describe the propagation of seismic waves, 
which establishes the relationship between the pressure and the velocity as
\begin{equation}
    \label{eq:acoustic-wave-equation}
   -(\nabla^2+\frac{\omega^2}{{v^{2}(\mathbf{x})}}) u(\mathbf{x}_{s},\mathbf{x},\omega) = Q(\mathbf{x}_{s},\mathbf{x},\omega)\delta(\mathbf{x}-\mathbf{x}_{s}),
\end{equation}
where 
\begin{align}
\omega=2\pi f,\quad {\nabla ^2} = \frac{{{\partial ^2}}}{{\partial {x^2}}} + \frac{{{\partial ^2}}}{{\partial {z^2}}},
\end{align}
where $v({\mathbf{x}})$ is the velocity at the point ${\mathbf{x}}=(x,z)$, 
$\mathbf{x}_{s}=(x_{s},z_{s})$ is the source location,
${\nabla ^2}$ denotes the Laplace operator, $u$ denotes the complex-valued pressure, $\omega$ and $f$ are angular frequency and frequency, respectively. $Q$ is the source function and $\delta$ is the Dirac Delta function.

Based on \eqref{eq:acoustic-wave-equation}, we can compute the simulated pressure fields at the receiver locations, which is formulated as  
\begin{equation} \label{forward-problem}
  \mathbf{d}=\mathbf{m}(\mathbf{v};\mathbf{x}_{s},f)+\bm{\epsilon}, 
\end{equation}
where $\mathbf{v}=\{v(\mathbf{x}^{(i)})\}^{L}_{i=1}$ denotes the velocity fields over a pre-defined mesh grids $\{\mathbf{x}^{(i)}\}^{L}_{i=1}$,  $L$ is the total number of mesh grid points. $\mathbf{d} \in \mathbb{C}^{M}$ is the corresponding observed pressures, with $M$ denoting the total number of receivers.  
Given a source location $\mathbf{x}_{s}$ and frequency $f$, the forward modeling operator $\mathbf{m}(\cdot) \in \mathbb{C}^{M}$ maps the velocity fields $\mathbf{v}$ to the simulated pressures.
$\bm{\epsilon} \in \mathbb{C}^M$ is the observation noise.  

FWI is commonly formulated as an optimization problem, where the goal is to infer the velocity fields by minimizing the misfit between the observed pressures $\mathbf{d}\in \mathbb{C}^M$ and the simulated pressures. The objective function can be expressed as follows: 

\begin{equation}  \label{eq:minimization-function}
\mathcal{L}(\mathbf{v})=\frac{1}{2}\sum^{I}_{i=1}\sum^{K}_{k=1}\left\Vert\mathbf{d}^{(i,k)}-\mathbf{m}^{(i,k)}\right\Vert^{2}_{2},
\end{equation}
where 
\begin{equation}
    \label{eq:forward-modeling}
    \mathbf{m}^{(i,k)}=\mathbf{m}(\mathbf{v};\mathbf{x}^{(i)}_{s},f^{(k)}),
\end{equation}
where $\{\mathbf{d}^{(i,k)}\}^{I,K}_{i=1,k=1}$ and $\{\mathbf{m}^{(i,k)}\}^{I.K}_{i=1,k=1}$ are obtained with $I$ sources and $K$ frequencies, respectively. 

Several optimization methods can be used to solve \eqref{eq:minimization-function}, involving steepest-descent, Gauss-Newton and L-BFGS methods. However, they often ignore to quantify the uncertainty about the inferred results. Therefore, we aim to devise an efficient algorithm to solve FWI problems and offer informative uncertainty estimations about the inferred velocity fields.

\section{Method}
\label{sec:method}
To quantify the uncertainty about the inverse results, Bayesian inference serves as a powerful approach that can be integrated into FWI. Bayesian inference offers a robust framework that integrates prior knowledge about parameters into the inference process, allowing for the assessment about the uncertainties of unknown parameters \cite{10.1093/gji/ggae329}. Various Bayesian inference methods can be employed to tackle the FWI problem, including Markov Chain Monte Carlo (MCMC) and Hamiltonian Monte Carlo (HMC). 
However, MCMC methods are usually computationally demanding, especially for high-dimensional inverse problems. Additionally, HMC requires  gradient information of the loss function with respect to the parameters, which can be computationally costly in high-dimensional scenarios. 

To enhance the computational efficiency in quantifying the uncertainty of  FWI problems and provide reasonable uncertainty estimates about the inversion results, we introduce ensemble Kalman inversion (EKI) algorithm as an efficient uncertainty quantification method for FWI framework. 

\subsection{Ensemble Kalman inversion} \label{sec:EKI}
We first provide a brief overview of ensemble Kalman inversion (EKI) algorithm. The EKI algorithm stems from ensemble Kalman filter (EnKF)  proposed by Evensen \cite{https://doi.org/10.1029/94JC00572}, which is a Bayesian inference method used to solve inverse problems. Several variants of the EKI have been developed in the literature \cite{pensoneault2023efficient,Iglesias_2016,Iglesias_2018,chada2022convergence,doi:10.1137/19M1242331}.
Here, we adopt the version of the EKI method described in \cite{chada2022convergence}.

In this study,  we aim to infer the subsurface velocity fields $\mathbf{v}$ using EKI algorithm. Notably, we use the unknown parameters $\bm{\xi}\in\mathbb{R}^{L}$ to parameterize the subsurface velocity fields $\mathbf{v}$, and the parameterization method is detailed in Section \ref{sec:GRF}.
The unknown parameters $\bm{\xi}$ is assumed to follow a prior distribution $p(\bm{\xi})$ and $\mathbf{G}(\mathbf{v})$ is the measurement operator. Consequently, the relationship between the measurement data $\mathbf{y}\in \mathbb{R}^{D}$ and the unknown parameters $\bm{\xi}$ is written as follows:
\begin{equation}
    \mathbf{y}=\mathbf{G}(\mathbf{v})+\bm{\eta}, \quad\mathbf{v}=\{\tilde{v}(\mathbf{x}^{(i)};\bm{\xi})\}^{L}_{i=1},
\end{equation}
where $\tilde{v}$ is the inferred velocity at point $\mathbf{x}^{(i)}$, $L$ is the dimensionality of the unknown parameters, $D$ is the dimensionality of the measurement data vector, $\bm{\eta} \in \mathbb{R}^{D}$ represents the measurement noise vector that follows a Gaussian distribution with zero mean and covariance $\bm{\Xi}\in\mathbb{R}^{D\times D}$. According to the Bayes' theorem, the posterior density function of the unknown parameters $\bm{\xi}$ is defined as
\begin{equation} \label{bayes-rules-1}
 p(\bm{\xi}|\mathbf{y})\propto p(\bm{\xi})p(\mathbf{y}|\bm{\xi}),
\end{equation}
where the likelihood function is defined as
\begin{equation} \label{bayes-rules-2}
 p(\mathbf{y}|\bm{\xi})\propto \exp\left(-\frac{1}{2}\left\Vert{\bm{\Xi}^{-1/2}}(\mathbf{y}- \mathbf{G}(\mathbf{v}))\right\Vert^{2}_{2}\right).
\end{equation}

To infer \eqref{bayes-rules-1}, several Bayesian inference methods including MCMC and HMC are often computationally expensive.
In contrast, the EKI algorithm provides an approximate solution for \eqref{bayes-rules-1} with several advantages. EKI avoids explicit gradient calculations, which is suitable for non-differentiable models. Meanwhile, it uses the sample statistics of ensemble members to implicitly approximate Hessian and gradient information \cite{Zhang_Xiao_Luo_He_2022}, thereby reducing computational complexity and storage requirements. 

The EKI method can be framed as state-space models \cite{PENSONEAULT2025113670}, and the state-space model formulation  corresponding to the original Bayesian inverse problem is given by
\begin{equation}
\label{eq:state-space-model}
\left\{
\begin{aligned}
\mathbf{y}_n &= \mathbf{G}(\mathbf{v}_{n}) + \bm{\eta}_n, \\
\bm{\eta}_{n} &\sim \mathcal{N}(\mathbf{0}, \bm{\Xi}),
\end{aligned}
\right.
\end{equation}

Based on \eqref{eq:state-space-model}, EKI utilizes the principles of the ensemble Kalman filter to address parameter estimation problems. During the initialization of EKI, an ensemble of $J$ initial parameters $\{\bm{\xi}_0^{(j)}\}_{j=1}^J$ is generated. At the $n$-th iteration, EKI updates the parameters using the Kalman update equation as follows: 
\begin{align} 
\label{eq:EKI-method}
\bm{\xi}^{(j)}_{n+1}=&\bm{\xi}^{(j)}_{n}+ \mathbf{C}^{\bm{\xi}\mathbf{G}}_{n}(\mathbf{C}^{\mathbf{G}\mathbf{G}}_{n}+h^{-1}\bm{\Xi})^{-1}\nonumber \\
&(\mathbf{y}-\bm{\eta}^{(j)}_{n}-\mathbf{G}(\mathbf{v}_{n}^{(j)})),
\end{align}
where $j=1,2,...,J$, $h$ represents the stepsize of the algorithm. The sample covariance matrices $\mathbf{C}^{\bm{\xi}\mathbf{G}}_{n}$ and $\mathbf{C}^{\mathbf{G}\mathbf{G}}_{n}$, the ensemble mean of parameters $\bar{\bm{\xi}}_{n}$ and the ensemble mean of corresponding measurement operator $\bar{\mathbf{G}}_{n}$ are defined as 
\begin{align}
\label{eq:EKI-covariance-1}
\mathbf{C}^{\bm{\xi}\mathbf{G}}_{n} &= \frac{1}{J-1}\sum^{J}_{j=1}(\bm{\xi}^{(j)}_{n}-\bar{\bm{\xi}}_{n})(\mathbf{G}(\mathbf{v}^{(j)}_{n})-\bar{\mathbf{G}}_{n})^{\top},\\
\label{eq:EKI-covariance-2}
\mathbf{C}^{\mathbf{G}\mathbf{G}}_{n} &= \frac{1}{J-1}\sum^{J}_{j=1}(\mathbf{G}(\mathbf{v}^{(j)}_{n})-\bar{\mathbf{G}}_{n})(\mathbf{G}(\mathbf{v}^{(j)}_{n})-\bar{\mathbf{G}}_{n})^{\top},\\
\label{eq:EKI-covariance-3}
\bar{\bm{\xi}}_{n}&=\frac{1}{J}\sum^{J}_{j=1}\bm{\xi}^{(j)}_{n}, \quad 
\bar{\mathbf{G}}_{n}=\frac{1}{J}\sum^{J}_{j=1}\mathbf{G}(\mathbf{v}^{(j)}_{n}).
\end{align} 
The ensemble of $J$ parameters $\{\bm{\xi}_{n}^{(j)}\}^{J}_{j=1}$ are iteratively updated using EKI algorithm as defined in \eqref{eq:EKI-method} until it satisfies the stopping criterion. The pseudocode of the EKI algorithm is summarized in Algorithm~\ref{EKI-algorithm-1}.
\begin{algorithm}
  
    \caption{EKI algorithm}
    \label{EKI-algorithm-1}
    
    \KwIn{1)\ initial parameters  \{$\bm{\xi}^{(j)}_{0}\}^{J}_{j=1}$,\newline
     2)\ the measurement data vector $\mathbf{y}$.}
    \KwOut{the inferred parameters \{$\bm{\xi}^{(j)}_{N}\}^{J}_{j=1}$.}
    \textbf{Set} $n\leftarrow 0$ \\
    \While{$n\leq N$}{
    1)\ solving the forward problem:\\
    \hspace*{4em}$\{\mathbf{G}(\mathbf{v}^{(j)}_{n})\}^{J}_{j=1}$.\\
    2)\ generating noises:\\
    \hspace*{4em}$\{\bm{\eta}^{(j)}_{n}\sim \mathcal{N}(0, \bm{\Xi})\}^{J}_{j=1}$.\\
    3)\ updating EKI:\\
    \hspace*{4em}$\bm{\xi}^{(j)}_{n+1}=\bm{\xi}^{(j)}_{n}+\mathbf{C}^{\bm{\xi} \mathbf{G}}_{n}(\mathbf{C}^{\mathbf{G}\mathbf{G}}_{n}+h^{-1}\bm{\Xi})^{-1}(\mathbf{y}-\bm{\eta}^{(j)}_{n}-\mathbf{G}(\mathbf{v}^{(j)}_{n}))$,\\
    \hspace*{1em}where\\ 
    \hspace*{4em}$\mathbf{C}^{\mathbf{G}\mathbf{G}}_{n}=\displaystyle\frac{1}{J-1}\sum^{J}_{j=1}{(\mathbf{G}(\mathbf{v}^{(j)}_{n})-\bar{\mathbf{G}}_{n})(\mathbf{G}(\mathbf{v}^{(j)}_{n})-\bar{\mathbf{G}}_{n})^{\top}}$,\\
    \hspace*{4em}$\mathbf{C}^{\bm{\xi} \mathbf{G}}_{n}=\displaystyle\frac{1}{J-1}\sum^{J}_{j=1}{(\bm{\xi}^{(j)}_{n}-\bar{\bm{\xi}}_{n})(\mathbf{G}(\mathbf{v}^{(j)}_{n})-\bar{\mathbf{G}}_{n})^{\top}}$,\\
    \hspace*{4em}$\bar{\bm{\xi}}_{n}=\displaystyle\frac{1}{J}\sum^{J}_{j=1}{\bm{\xi}^{(j)}_{n}},\quad\bar{\mathbf{G}}_{n}=\displaystyle\frac{1}{J}\sum^{J}_{j=1}{\mathbf{G}(\mathbf{v}^{(j)}_{n})}$.\\
    4) stopping criterion:\\
    \If {\text{convergence}}{
        $N\leftarrow n$ \\
        \textbf{Break}  
    }
    $n\leftarrow n+1$}
\end{algorithm} 

\subsection{Gaussian Random Fields 
} \label{sec:GRF}
Constructing an appropriately designed initial ensemble of parameters $\{\bm{\xi}^{j}_{0}\}_{j=1}^{J}$ is beneficial for EKI algorithm, as it ensures adequate representation of the parameter space.
In this study, we utilize GRFs to generate the initial ensemble of parameters for EKI. 
By incorporating prior knowledge of the length-scale and smoothness, 
the initial ensemble generated by GRFs captures the spatial correlation characteristics of the parameters \cite{4847c1a8-518e-3c37-8c1a-780bcc44d236}.

We use GRFs with Whittle-Matérn correlation function \cite{10.5555/1162254, Lassi} to define the spatial correlation structure of the parameters. The Wittle-Matérn covariance function is defined as
\begin{equation} 
    \label{eq:matern}
    C(\mathbf{x}, \hat{\mathbf{x}})=\tau^{2}\frac{2^{1-\nu}}{\Gamma(\nu)}\left(\frac{\|\mathbf{x}-\hat{\mathbf{x}}\|_{2}}{\lambda}\right)^{\nu}K_{\nu}\left(\frac{\|\mathbf{x}-\hat{\mathbf{x}}\|_{2}}{\lambda}\right),
\end{equation}
 where $\lambda>0$ represents the length-scale, $\tau$ denotes the amplitude, $\nu>0$ is the smoothness, $\Gamma$ is the Gamma function, and $K_{\nu}$ denotes the modified Bessel function of the second kind with order of $\nu$. 
 
The function $C(\mathbf{x}, \hat{\mathbf{x}})$ characterizes the spatial correlation between the points $\mathbf{x}$ and $\hat{\mathbf{x}}$.    As
mentioned in Section \ref{sec:problem}, we employ a pre-defined mesh grid  $\{\mathbf{x}^{(i)}\}^{L}_{i=1}$ to discrete the velocity fields $\mathbf{v}=\{v(\mathbf{x}^{(i)})\}^{L}_{i=1}$, where $v(\mathbf{x}^{(i)})$ represents the velocity at the point $\mathbf{x}^{(i)}$.
Consequently, the spatial correlations between all points in the pre-defined mesh can be fully described by the covariance matrix $\bm{\Phi}$ as
 \begin{equation} \label{GRFs-covariance}
\bm{\Phi}=\begin{bmatrix}
        C(\mathbf{x}^{(1)}, \mathbf{x}^{(1)}) & \cdots &  C(\mathbf{x}^{(1)}, \mathbf{x}^{(L)}) \\
        \vdots & \ddots & \vdots \\    
        C(\mathbf{x}^{(L)}, \mathbf{x}^{(1)}) & \cdots & C(\mathbf{x}^{(L)}, \mathbf{x}^{(L)})
        \end{bmatrix}.
\end{equation}

Based on \eqref{GRFs-covariance},  the initial ensemble of EKI can be generated by  sampling from a zero-mean Gaussian distribution with covariance matrix  $\bm{\Phi}$, as described below:
\begin{equation}
    \label{eq:GRFs-samples}
    \bm{\xi}_{0}^{(j)}\sim\mathcal{N}(\mathbf{0},\bm{\Phi}),
\end{equation}
where $j=1,2,3,\dots,J$. Due to the large dimensionality of the covariance matrix $\bm{\Phi}$, generating the parameters through directly sampling from \eqref{eq:GRFs-samples} can be computationally expensive. To address this issue, several efficient methods can be used to simulate GRFs, involving spectral \cite{liu2019advances}, circulant embedding \cite{doi:10.1137/17M1149730}, Karhunen-Loève expansion \cite{BETZ2014109}, moving average \cite{ravalec2000fft} methods. In this work, we use the Karhunen-Loève expansion method from the GRFs software package \cite{robbe2023gaussianrandomfields} to generate the parameters.

\begin{figure}[!ht]
    \centering
    \includegraphics[width=0.6\linewidth]{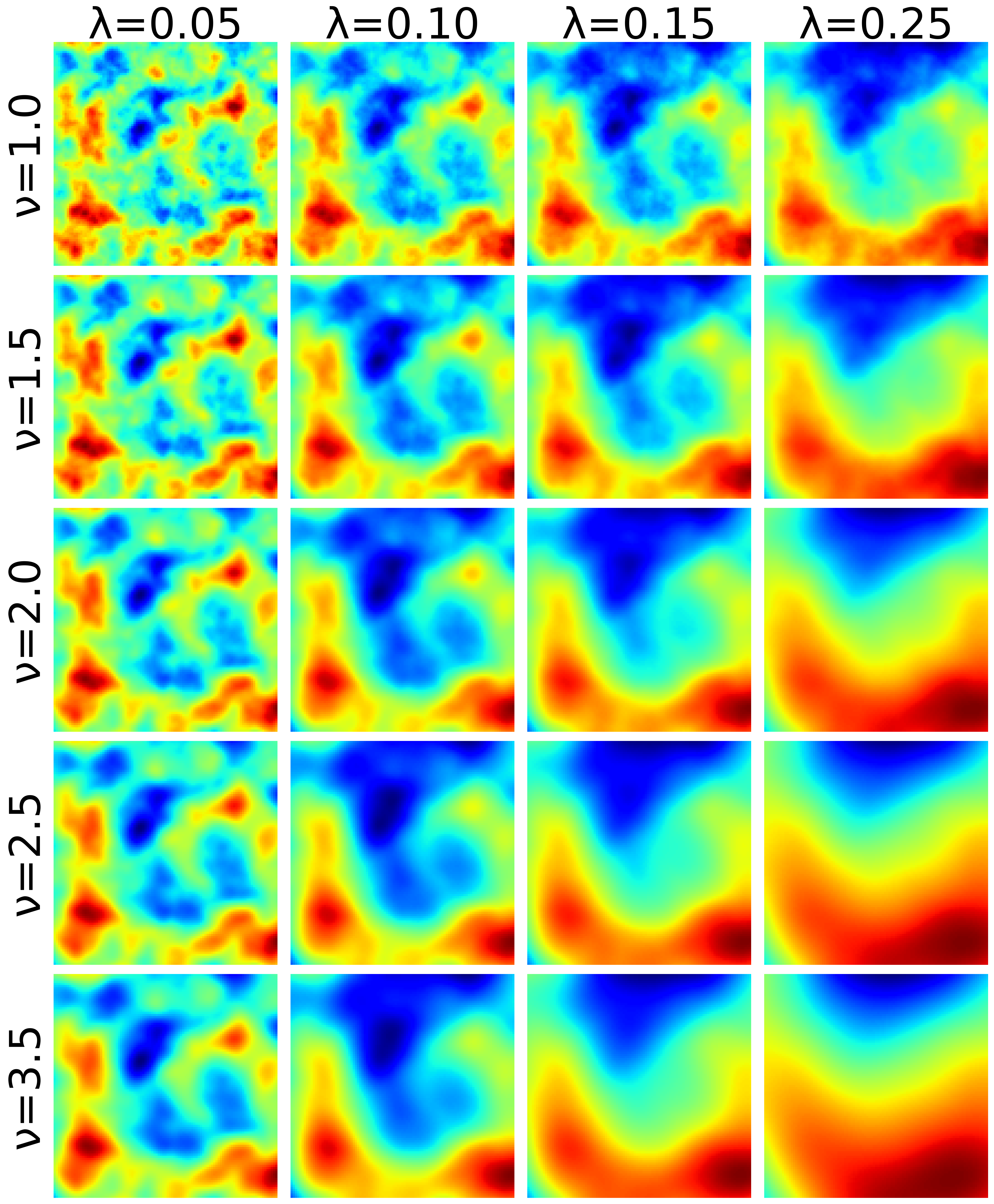}
    \caption{Sampled Gaussian Random Fields with different length-scale $\lambda$ and smoothness $\nu$.}
    \label{fig:sampled-GRFs}
\end{figure}

Notably, the hyperparameters including length-scale $\lambda$ and smoothness $\nu$ in \eqref{eq:matern} should be carefully chosen as they determine the spatial scale and smoothness of generated parameters, respectively. Fig.~\ref{fig:sampled-GRFs} illustrates the parameters sampled from GRFs with $\lambda=0.05,0.10,0.15,0.25$ and $\nu=1.0,1.5,2.0,2.5,3.5$. We observe that as $\nu$ varies from $1.0$ to $3.5$, the parameter fields gradually behaves from rough to smooth. When $\lambda$ varies from $0.05$ to $0.25$,   both the spatial scale and smoothness of the parameter fields exhibit significant variations. Based on effects of $\lambda$ and $\nu$ on the parameter fields, we investigate the impact of the initial parameters generated by GRFs when $\lambda=0.05,0.10,0.15,0.25$ on the inferred velocity fields by EKI. For the hyperparameter smoothness, we set $\nu=2.0$.

When inferring subsurface velocity fields, the upper and lower bounds of velocity values can be obtained from the prior knowledge. 
Based on the upper and lower bounds of velocity values, the inferred velocity fields falls into a physically realistic range by mapping the parameters $\bm{\xi}$ to the velocity $\tilde{v}(\cdot)$ using the following equation:
 \begin{equation}
 \label{eq:rescale}
    \tilde{v}(\mathbf{x}^{(i)};\bm{\xi})=\frac{1}{1+\exp(-\xi^{(i)})}\times (v_{\max}-v_{\min}) + v_{\min},
\end{equation}
where $\bm{\xi}=\{\xi^{(i)}\}^{L}_{i=1}$, and $\xi^{(i)}$ represents the parameter at point $\mathbf{x}^{(i)}$. $v_{\max}$ and $v_{\min}$ represent the upper and lower limits of the velocity fields.

\subsection{EKI-GRFs-FWI}
 Next, we integrate EKI method with FWI framework, which can efficiently infer the velocity fields $\mathbf{v}$ and provide uncertainty quantification for the inferred results.
As mentioned in Section~\ref{sec:GRF}, we sample by GRFs to generate an initial ensemble for EKI, and the initial ensemble consists of $J$ parameters $\{\bm{\xi}^{(j)}_{0}\sim\mathcal{N}(\mathbf{0},\bm{\Phi})\}^{J}_{j=1}$. 
During each iteration in \eqref{eq:EKI-method}, the parameters $\{\bm{\xi}^{(j)}_{n}\}^{J}_{j=1}$ are mapped to the velocity fields using \eqref{eq:rescale}, and the resulting velocity fields can be expressed as
\begin{equation}
    \mathbf{v}^{(j)}_{n}=\left\{\tilde{v} (\mathbf{x}^{(i)};\bm{\xi}^{(j)}_{n})\right\}^{L}_{i=1}, \quad
    j=1,2,3,\dots,J.
\end{equation} 
The resulting velocity fields  are used for the measurement operator $\mathbf{G}(\mathbf{v}^{(j)}_{n})$ at each update of EKI method in \eqref{eq:EKI-method}.
To integrate EKI method with FWI framework, we shall redefine the measurement operator $\mathbf{G}$, which 
incorporates a set of simulated pressure fields $\{\mathbf{m}^{(i,k)}\}^{I,K}_{i=1,k=1}$ computed at all $I$ sources and  $K$ frequencies in \eqref{eq:forward-modeling}. Since each $\mathbf{m}^{(i,k)}$ is a complex-valued vector, the parameters $\bm{\xi}^{(j)}_{n+1}$ also is a complex-valued vector after each update of EKI methd in \eqref{eq:EKI-method}. However, each parameter $\bm{\xi}$ should be the real-valued vector in \eqref{eq:rescale}. 
To ensure that the parameters 
$\bm{\xi}^{(j)}_{n}$ remains real-valued after each update of the EKI methd, we concatenate the real part and imaginary part of each complex-valued vector $\mathbf{m}^{(i,k)}$ to a real vector. 
Consequently, the measurement operator $\mathbf{G}(\cdot) \in \mathbb{R}^{D}$ ($D=2 \times I \times K \times M$) can be expressed as follows:
\begin{align}
    \mathbf{G}(\mathbf{v})=
    &\left[\mathbf{m}^{(1,1)^{\top}}_{\Re},
    \mathbf{m}^{(1,2)^{\top}}_{\Re},\dots, \mathbf{m}^{(I,K)^{\top}}_{\Re},\right. \nonumber\\ 
    &\ \ \mathbf{m}^{(1,1)^{\top}}_{\Im},  \left. \mathbf{m}^{(1,2)^{\top}}_{\Im},\dots, 
    \mathbf{m}^{(I,K)^{\top}}_{\Im}
    \right]^{\top},
\end{align}
where 
\begin{equation}
    \mathbf{m}^{(i,k)}_{\Re}=\Re\left(\mathbf{m}^{(i,k)}\right),\quad
    \mathbf{m}^{(i,k)}_{\Im}=\Im\left(\mathbf{m}^{(i,k)}\right),
\end{equation}
 where $\mathbf{m}^{(i,k)} \in \mathbb{C}^{M}$, and the symbols $\Re$ and $\Im$ are used to extract the real and imaginary parts of a complex-valued vector, respectively. Similarly, the measurement data vector $\mathbf{y} \in \mathbb{R}^{D}$ is written as
\begin{align}
    \mathbf{y}=
    &\left[\mathbf{d}^{(1,1)^{\top}}_{\Re},
    \mathbf{d}^{(1,2)^{\top}}_{\Re},\dots, \mathbf{d}^{(I,K)^{\top}}_{\Re},\right. \nonumber\\ 
    &\ \ \mathbf{d}^{(1,1)^{\top}}_{\Im},  \left. \mathbf{d}^{(1,2)^{\top}}_{\Im},\dots, 
    \mathbf{d}^{(I,K)^{\top}}_{\Im}
    \right]^{\top},
\end{align}
where 
\begin{equation}
    \mathbf{d}^{(i,k)}_{\Re}=\Re\left(\mathbf{d}^{(i,k)}\right),\quad
    \mathbf{d}^{(i,k)}_{\Im}=\Im\left(\mathbf{d}^{(i,k)}\right),
\end{equation}
where $\{\mathbf{d}^{(i,k)}\}^{I,K}_{i=1,k=1}$ are the observed pressures at the receiver locations obtained with all $I$ sources and  $K$ frequencies in \eqref{eq:minimization-function}, and $\mathbf{d}^{(i,k)} \in \mathbb{C}^{M}$, $M$ is the total number of the receivers.  

\subsection{Mini-batch strategy}
The dimensionalities of the sample covariance matrices $\mathbf{C}^{\bm{\xi}\mathbf{G}}_{n}$ in \eqref{eq:EKI-covariance-1} and $\mathbf{C}^{\mathbf{G}\mathbf{G}}_{n}$ in \eqref{eq:EKI-covariance-2} are related to the dimensionality of the measurement data $\mathbf{y}$.
When the dimensionality of $\mathbf{y}$ is large, computing high-dimensional matrix $\mathbf{C}^{\bm{\xi}\mathbf{G}}_{n}$  and $\mathbf{C}^{\mathbf{G}\mathbf{G}}_{n}$ leads to high computational costs and excessive memory usage. To address this issue, we propose a strategy that reduces the dimensionality of the $\mathbf{C}^{\bm{\xi}\mathbf{G}}_{n}$ and $\mathbf{C}^{\mathbf{G}\mathbf{G}}_{n}$ by splitting the measurement data vector into several mini-batches based on the number of frequencies $K$. 
Specifically, we divide the measurement data vector $\mathbf{y}$ into $K$ sub-measurement data vector, and each sub-measurement data vector is written as 
\begin{align}
    \mathbf{y}^{(k)}=
    &\left[\mathbf{d}^{(1,k)^{\top}}_{\Re},
    \mathbf{d}^{(2,k)^{\top}}_{\Re},\dots, \mathbf{d}^{(I,k)^{\top}}_{\Re},\right. \nonumber\\ 
    &\ \ \mathbf{d}^{(1,k)^{\top}}_{\Im},  \left. \mathbf{d}^{(2,k)^{\top}}_{\Im},\dots, 
    \mathbf{d}^{(I,k)^{\top}}_{\Im}
    \right]^{\top},
\end{align}
where the sub-measurement data vector $\mathbf{y}^{(k)}\in \mathbb{R}^{H}$ ($H=2 \times I \times M$) represents the observed pressures at receiver locations obtained with all $I$ sources at frequency $f^{(k)}$, and $k=1,2,3,\dots,K$. Correspondingly, the sub-measurement operator is formulated as
\begin{align}
     \mathbf{G}(\mathbf{v}, k)=
    &\left[\mathbf{m}^{(1,k)^{\top}}_{\Re},
    \mathbf{m}^{(2,k)^{\top}}_{\Re},\dots, \mathbf{m}^{(I,k)^{\top}}_{\Re},\right. \nonumber\\ 
    &\ \ \mathbf{m}^{(1,k)^{\top}}_{\Im},  \left. \mathbf{m}^{(2,k)^{\top}}_{\Im},\dots, 
    \mathbf{m}^{(I,k)^{\top}}_{\Im}
    \right]^{\top},
\end{align}
where the sub-measurement operator $\mathbf{G}(\mathbf{v},k)\in \mathbb{R}^{H}$ computes the simulated pressures at receiver locations with all $I$ sources at frequency $f^{(k)}$. 

 Assuming that each sub-measurement data vector is corrupted by Gaussian random noise, the measurement noise $\bm{\eta}\sim \mathcal{N}(\mathbf{0},\bm{\Xi})$ in \eqref{eq:EKI-method} is redefined as $\hat{\bm{\eta}} \in \mathbb{R}^{H}$, and its covariance matrix $\hat{\bm{\Xi}}\in \mathbb{R}^{H\times H}$ is defined as follows:
\begin{equation}
\label{eq:measurement-noise}
\left\{
\begin{aligned}
    \hat{\bm{\Xi}} &= \begin{bmatrix}
        \sigma_{\bm{\eta}_{\Re}}^{2}\mathbf{I} & \mathbf{0}\\
        \mathbf{0} & \sigma_{\bm{\eta}_{\Im}}^{2}\mathbf{I}
    \end{bmatrix},\\
    \sigma_{\bm{\eta}_{\Re}} &= \alpha\frac{1}{I}\frac{1}{K}\frac{1}{M}\sum^{I}_{i=1}\sum^{K}_{k=1}\left\Vert\mathbf{m}_{\Re}^{(i,k)}\right\Vert_{1},\\
    \sigma_{\bm{\eta}_{\Im}} &= \alpha\frac{1}{I}\frac{1}{K}\frac{1}{M}\sum^{I}_{i=1}\sum^{K}_{k=1}\left\Vert\mathbf{m}_{\Im}^{(i,k)}\right\Vert_{1},\\
\end{aligned}
\right.
\end{equation}
where $\sigma_{\bm{\eta}_{\Re}}$ and $\sigma_{\bm{\eta}_{\Im}}$ are the standard deviations of the measurement noise, $\mathbf{I}$ is the identity matrix, $\alpha$ is the noise level, $\Vert\cdot\Vert_{1}$ denotes the $l_{1}$ norm of a vector.

To apply the mini-batch strategy for EKI algorithm, $\mathbf{G}(\mathbf{v})$, $\mathbf{y}$, $\bm{\eta}$ and $\bm{\Xi}$ in \eqref{eq:EKI-method}, \eqref{eq:EKI-covariance-1}, \eqref{eq:EKI-covariance-2} and \eqref{eq:EKI-covariance-3} can be replaced using $\mathbf{G}(\mathbf{v},k)$, $\mathbf{y}^{(k)}$, $\hat{\bm{\eta}}$ and $\hat{\bm{\Xi}}$, respectively.
Subsequently, we derive a mini-batch version of EKI algorithm combined with the GRFs to solve the FWI problems, referred to as EKI-GRFs-FWI algorithm. 

To ensure that EKI-GRFs-FWI algorithm terminates properly, we adopt a modified version of the stopping criterion introduced in \cite{10.1093/gji/ggae329}, which is defined as follows:
\begin{align}
    \label{eq:stopping-criterion-1}
    &D_{n} =\frac{1}{2}\sum^{I}_{i=1}\sum^{K}_{k=1}\left\Vert\mathbf{d}^{(i,k)}-\mathbf{m}^{(i,k)}\right\Vert^{2}_{2},\\
    \label{eq:stopping-criterion-2}
    &\bar{D} = \frac{1}{W}\sum^{n}_{n-W}D_{n}, \\
    \label{eq:stopping-criterion-3}
    &\max_{m \in \{n-W, \dots, n\}} \frac{|D_{m} - \bar{D}|}{\bar{D}} < \rho,
\end{align}
where $D_{n}$ denotes the discrepancy metric at the $n$-th iteration, while $\bar{D}$ indicates the average of the discrepancy metric over a sliding window of length $W$.   If the relative change in $D_{n}$ within the window is less than a pre-defined threshold $\rho$, EKI-GRFs-FWI algorithm terminates. The pseudocode of EKI-GRFs-FWI algorithm is summarized in
Algorithm \ref{EKI-algorithm-2}.
\begin{algorithm}[!ht]
    \caption{The Mini-batch version of EKI-GRFs-FWI algorithm}
    \label{EKI-algorithm-2}    
    \KwIn{1)\ generating initial parameters  \{$\bm{\xi}^{(j)}_{0}\}^{J}_{j=1}$ by sampling from GRFs,\newline
     2)\ all sub-measurement data vectors $\{\mathbf{y}^{(k)}\}^{K}_{k=1}$.}
    \KwOut{the inferred parameters \{$\bm{\xi}^{(j)}_{N}\}^{J}_{j=1}$.}
    \textbf{Set} $n\leftarrow 0$, $k\leftarrow 1$ \\
    \While{$n\leq N$}{
    1)\ solving the forward problem:\\
    \hspace*{4em}$\{\mathbf{G}(\mathbf{v}^{(j)}_{n},k)\}^{J}_{j=1}$.\\
    2)\ generating noises:\\
    \hspace*{4em}$\{\hat{\bm{\eta}}^{(j)}_{n}\sim \mathcal{N}(0, \hat{\bm{\Xi}})\}^{J}_{j=1}$.\\
    3)\ updating EKI-GRFs-FWI:\\
    \hspace*{4em}$\bm{\xi}^{(j)}_{n+1}=\bm{\xi}^{(j)}_{n}+\mathbf{C}^{\bm{\xi} \mathbf{G}}_{n}(\mathbf{C}^{\mathbf{G}\mathbf{G}}_{n}+h^{-1}\mathbf{\hat{\bm{\Xi}}})^{-1}(\mathbf{y}^{(k)}-\hat{\bm{\eta}}^{(j)}_{n}-\mathbf{G}(\mathbf{v}^{(j)}_{n},k))$,\\
    \hspace*{1.3em}where\\ 
    \hspace*{4em}$\mathbf{C}^{\mathbf{G}\mathbf{G}}_{n}=\displaystyle\frac{1}{J-1}\sum^{J}_{j=1}{(\mathbf{G}(\mathbf{v}^{(j)}_{n},k)-\bar{\mathbf{G}}_{n})(\mathbf{G}(\mathbf{v}^{(j)}_{n},k)-\bar{\mathbf{G}}_{n})^{\top}}$,\\
    \hspace*{4em}$\mathbf{C}^{\bm{\xi} \mathbf{G}}_{n}=\displaystyle\frac{1}{J-1}\sum^{J}_{j=1}{(\bm{\xi}^{(j)}_{n}-\bar{\bm{\xi}}_{n})(\mathbf{G}(\mathbf{v}^{(j)}_{n},k)-\bar{\mathbf{G}}_{n})^{\top}}$,\\
    \hspace*{4em}$\bar{\bm{\xi}}_{n}=\displaystyle\frac{1}{J}\sum^{J}_{j=1}{\bm{\xi}^{(j)}_{n}},\quad \bar{\mathbf{G}}_{n}=\displaystyle\frac{1}{J}\sum^{J}_{j=1}{\mathbf{G}(\mathbf{v}^{(j)}_{n},k)}$.\\
    4) stopping criterion:\\
    \If {$n>W$ \rm{\textbf{and}} $\max\limits_{m\in \{n-W,...,n\}}\frac{|D_{m}-\bar{D}|}{\bar{D}}<\rho$}{
        $N\leftarrow n$ \\
        \textbf{Break}  
    }
    $n\leftarrow n+1$\\
    5) updating frequency:\\
    $k\leftarrow k+1$\\
    \If {$k>K$}{
        $k\leftarrow 1$
    }
    }
\end{algorithm}

After we obtain all $J$ inferred velocity fields $\{\tilde{\mathbf{v}}=\{\tilde{v}(\mathbf{x}^{(i)},\bm{\xi}^{(j)}_{N})\}^{L}_{i=1}\}^{J}_{j=1}$,  we can calculate the posterior mean of the $J$ inferred velocity fields as the estimated velocity fields $\bar{\mathbf{v}}_{N}$, which can be formulated as follows:
\begin{align} \label{eq:posterior-v}
\left\{
\begin{aligned}
\bar{\mathbf{v}}_{N} &= \{\bar{v}(\mathbf{x}^{(i)})\}_{i=1}^{L}, \\
\bar{v}(\mathbf{x}^{(i)}) &= \frac{1}{J}\sum^{J}_{j=1}\tilde{v}(\mathbf{x}^{(i)};\bm{\xi}^{(j)}_{N}).
\end{aligned}
\right.
\end{align}

To quantify the uncertainty of the estimated velocity fields, we utilize the posterior standard deviation (STD) as a measure of uncertainty. A smaller STD indicates a greater confidence in the estimated velocities, whereas a larger STD indicates higher uncertainty in the velocity estimates. The STD $\mathbf{S}_{N}$ is defined as follows:
\begin{equation}
\label{eq:std-v}
\left\{
\begin{aligned}
    \mathbf{S}_{N} &= \left\{s(\mathbf{x}^{(i)})\right\}^{L}_{i=1}, \\
    s(\mathbf{x}^{(i)}) &= \sqrt{\frac{1}{J-1}\sum^{J}_{j=1}\left(\tilde{v}(\mathbf{x}^{(i)};\bm{\xi}^{(j)}_{N}) - \bar{v}(\mathbf{x}^{(i)})\right)^{2}}.
\end{aligned}
\right.
\end{equation}

\begin{figure*}[!ht]
    \centering
    \includegraphics[width=0.99\linewidth]{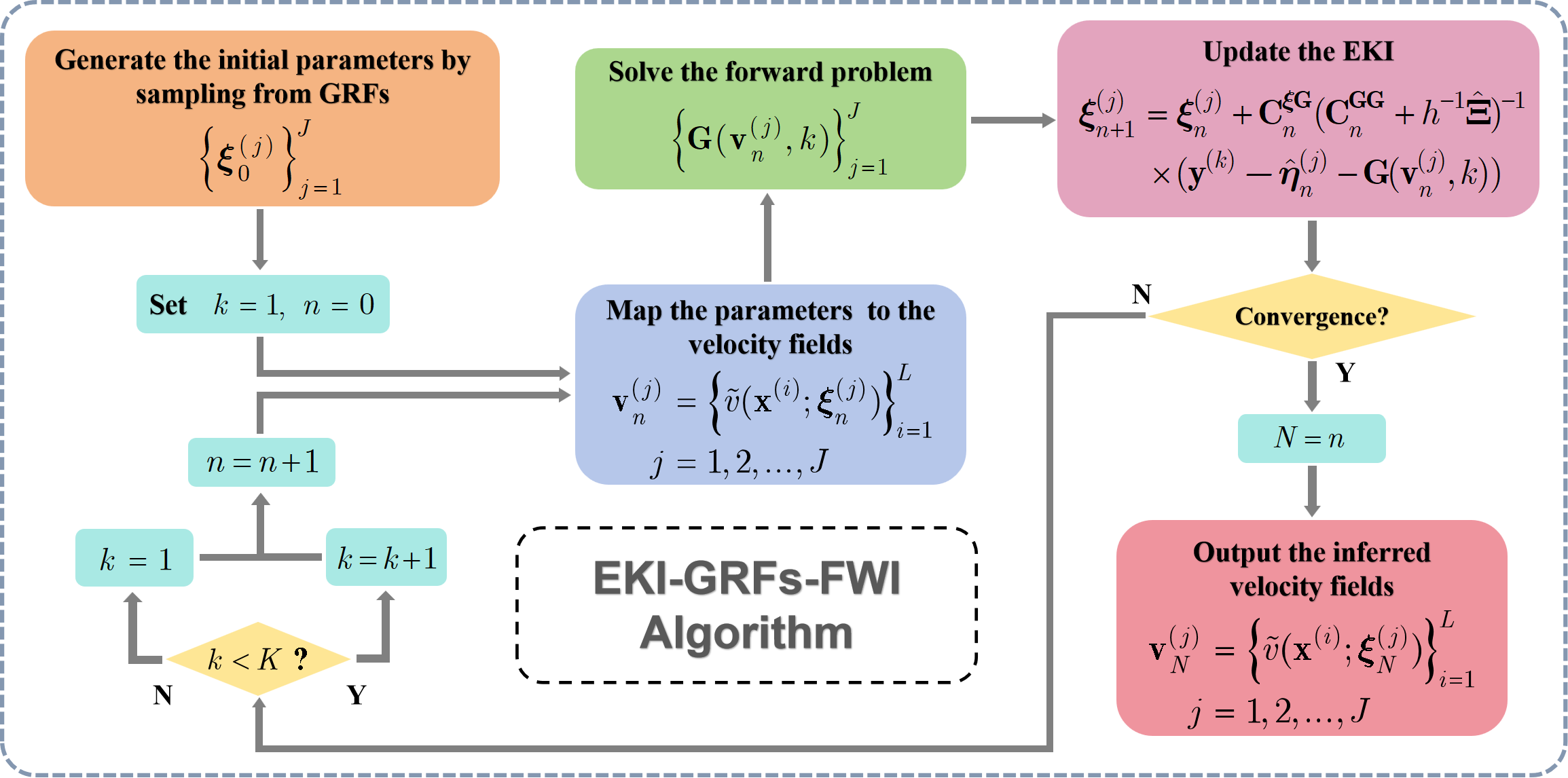}
    \caption{The workflow of the EKI-GRFs-FWI algorithm. $n$ represents the number of iterations and $k$ is the index of frequency $f^{(k)}$.}
    \label{fig:EKI-GRFs-FWI-framework}
\end{figure*}

Overall, the workflow of EKI-GRFs-FWI method is shown in Fig.~\ref{fig:EKI-GRFs-FWI-framework}.

\section{Result}
We will conduct a series of experiments involving several velocity models to illustrate the feasibility and effectiveness of the proposed method, which includes the inclusion, checkerboard, and overthrust models. 

{\bf Generation of measurement data}. 
We generate the noise-free 
observed pressure fields at all frequencies $\{f^{(k)}\}_{k=1}^{K}$ using finite difference method \cite{10.1190/geo2014-0085.1} that solves the frequency-domain acoustic wave equation. The source is represented by a Ricker wavelet with a dominant frequency of 10 Hz. 
Additionally, each sub-measurement data vector is corrupted by the Gaussian random noise $\bm{\phi}\sim\mathcal{N}(\mathbf{0},\hat{\bm{\Xi}})$ with the noise level $\alpha=0.05$ (5\%) in \eqref{eq:measurement-noise}.

{\bf Setup for EKI}. In this work,
the initial ensemble for EKI is generated by GRFs, where the 
smoothness $\nu=2.0$ and amplitude $\tau=1.0$ as in \eqref{eq:matern}. For the length-scale parameter, we select $\lambda=0.05, 0.1,0.15$ and $0.25$ to investigate the impact of initial parameters generated by GRFs with different length-scales. For the stopping criterion, we set $W=10$ and $\rho=0.1$ in  \eqref{eq:stopping-criterion-2} and \eqref{eq:stopping-criterion-3}.  

{\bf Parallelization}. EKI is highly parallelizable due to the independent computations performed on multiple ensemble members, which allows for parallelization on the threads as many as the 
ensemble members.
In this work, all experiments are implemented on Windows 64-bit with Intel i9-10900X CPU with 10 cores, and the code is parallelized on 10 threads.

To assess the computational efficiency and reliability of the uncertainty estimations by EKI-GRFs-FWI, a gradient-based MCMC method \cite{doi:10.1137/19M1284014, 10.1093/gji/ggab287} (\url{https://github.com/izzatum/Langevin_GJI_2020/blob/main/Sampler/samplerMALA.m}) is chosen as the reference solution. This method employs the Metropolis-adjusted Langevin algorithm (MALA) to guide the sampling process of MCMC.
To fairly compare their computational time, we run the numerical solver for MCMC method with 10 threads, where each thread runs the numerical solver at a specific frequency. 
For all experiments, the acquisition system and observed data used in the EKI-GRFs-FWI and MCMC methods are the same.
Additionally, the EKI-GRFs-FWI and MCMC methods are implemented in Matlab. 

{\bf Error metric}.  
We examine  the accuracy of the inferred velocity fields by EKI-GRFs-FWI and MCMC methods according to the relative error (RE): 
\begin{equation} \label{RE}
    {{\rm{RE}}} = \sqrt{\frac{\sum^{L}_{i=1}|\bar{v}(\mathbf{x}^{(i)})-v(\mathbf{x}^{(i)})|^{2}}{\sum^{L}_{i=1}|v(\mathbf{x}^{(i)})|^{2}}}, 
\end{equation}
where $\{\bar{v}(\mathbf{x}^{(i)})\}^{L}_{i=1}$ is the estimated velocity fields obtained by EKI-GRFs-FWI or MCMC method, and  $\{v(\mathbf{x}^{(i)})\}^{L}_{i=1}$ is the ground-truth velocity fields, and $L$ is the total number of the pre-defined mesh grid points.

\subsection{Inclusion model}
\begin{figure*}[!ht]
    \centering
    \includegraphics[width=0.9\linewidth]{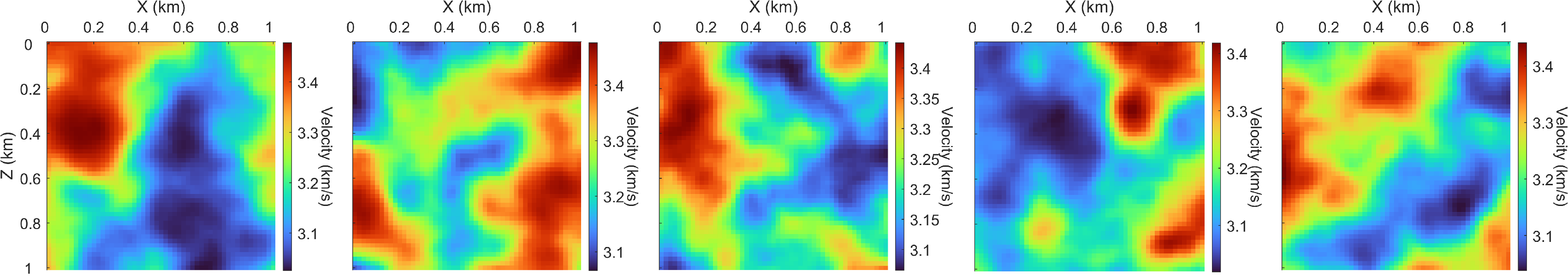}

    \caption{The initial velocity fields for inclusion model, which is randomly selected from the initial ensemble of velocity fields. GRFs is leveraged to generate the initial ensemble of parameters using \eqref{eq:GRFs-samples}, and then the initial ensemble of parameters is mapped to the initial ensemble of velocity fields using \eqref{eq:rescale}.}
    \label{fig:EKI-GRFs-FWI-Inclusion-ensemble-velocity-fields}
\end{figure*}

\begin{figure*}[!ht]
    \centering
    \includegraphics[width=0.99\linewidth]{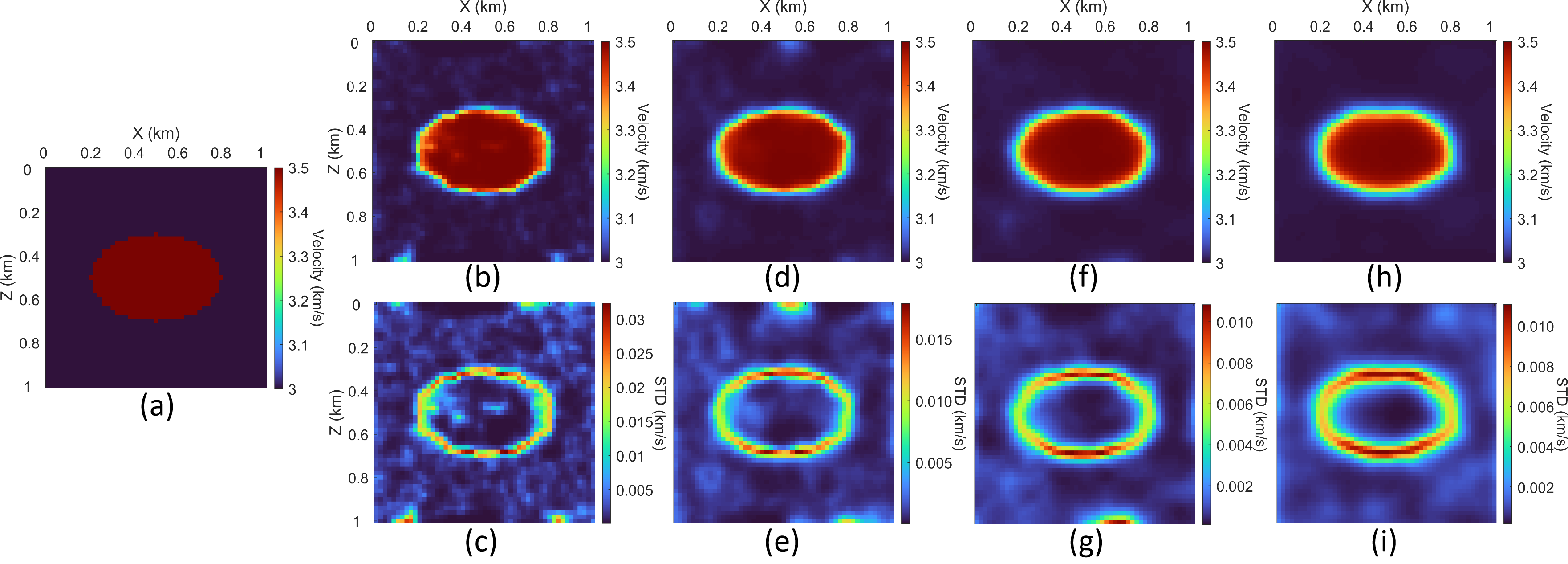}

    \caption{The result of EKI-GRFs-FWI with different length-scale $\lambda$ for inclusion model: (a) ground-truth model, (b) and (c) the estimated velocity fields and its posterior standard deviations when $\lambda=0.05$, (d) and (e) the estimated velocity fields and its posterior standard deviations when $\lambda=0.1$, (f) and (g) the estimated velocity fields and its posterior standard deviations when $\lambda=0.15$, (h) and (i) the estimated velocity fields and its posterior standard deviations when $\lambda=0.25$.  }
    \label{fig:EKI-GRFs-FWI-Inclusion-result}
\end{figure*}

We first test an inclusion model to verify the feasibility of EKI-GRFs-FWI, which is shown in Fig.~\ref{fig:EKI-GRFs-FWI-Inclusion-result} (a). The size of the velocity model is 1.0 km $\times$ 1.0 km and the grid size is 20 m, resulting in a mesh size of 51 $\times$ 51. The acquisition system employs a cross-well arrangement,  where 17 sources are evenly arranged along $X$ = 0 km, and 51 receivers are uniformly placed along $X$ = 1 km. We infer the velocity fields using the observed data with the frequencies $f^{(k)}=[3, 3.5, 4, 4.5, 5, 6, 7, 8, 9, 10]$ Hz.
For the EKI hyperparameters, the ensemble size $J$ is set to 500, and stepsize $h=0.5$ as in \eqref{eq:EKI-method}. To better understand the initial velocity fields for EKI, Fig.~\ref{fig:EKI-GRFs-FWI-Inclusion-ensemble-velocity-fields} shows five initial velocity fields by randomly selecting from the initial ensemble of velocity fields. GRFs first generate the initial ensemble of parameters using \eqref{eq:GRFs-samples}, and then the initial ensemble of parameters is mapped to the initial ensemble of velocity fields using \eqref{eq:rescale}.
\begin{table*}[!ht]
    \caption{The error metric of estimated velocity and computational time obtained by EKI-GRFs-FWI and MCMC.}
    \centering
    \begin{tabular}{c c c c c c }
    \toprule
       Model & Mesh size  & Method & ${\rm{RE}}$ & Computational time  & Iterations\\
     \midrule
     \multirow{5}{*}{Inclusion} & \multirow{5}{*}{51 $\times$ 51 (2601)}  & EKI-GRFs-FWI ($\lambda=$ 0.05)  & 0.0150 &  54.0 s & 40\\
     & & EKI-GRFs-FWI ($\lambda=$ 0.10)  &  0.0156 & 43.3 s & 32 \\
     & & EKI-GRFs-FWI ($\lambda=$ 0.15)  & 0.0165 &  42.1 s & 32 \\
     & & EKI-GRFs-FWI ($\lambda=$ 0.25)  & 0.0181 &   43.6 s & 32 \\
     & & MCMC (homogeneous initial model)  &  0.0146  &  21 min 16 s & 10000 \\
     \midrule
    \multirow{5}{*}{Checkerboard} & \multirow{5}{*}{97 $\times$ 33 (3201)}   & EKI-GRFs-FWI ($\lambda=$ 0.05)  & 0.0215 &  1 min 41 s & 32\\
     & & EKI-GRFs-FWI ($\lambda=$ 0.10)  & 0.0233 &  1 min 40 s & 32\\
     & & EKI-GRFs-FWI ($\lambda=$ 0.15)  & 0.0250 &  1 min 40 s & 32\\
     & & EKI-GRFs-FWI ($\lambda=$ 0.25)  & 0.0311 &  1 min 41 s & 32\\
     & & MCMC (homogeneous initial model)  &  0.0253  &  31 min 40 s & 10000 \\
       \midrule
      \multirow{6}{*}{Overthrust} & \multirow{6}{*}{201 $\times$ 46 (9246)}  & EKI-GRFs-FWI ($\lambda=$ 0.05)  & 0.0493 &  18 min 40 s & 31\\
     & & EKI-GRFs-FWI ($\lambda=$ 0.10)  & 0.0525 &  18 min 37 s & 31\\
     & & EKI-GRFs-FWI ($\lambda=$ 0.15)  & 0.0523 &  18 min 09 s & 30\\
     & & EKI-GRFs-FWI ($\lambda=$ 0.25)  & 0.0540 &  18 min 40 s & 31\\
     & & MCMC (homogeneous initial model)  & 0.1772  &  2 h 25 min & 20000 \\
     & & MCMC (smoothed initial model)  & 0.0491  &  2 h 25 min & 20000 \\
    \bottomrule
    \end{tabular}
     \label{table:error-metric-and-time-EKI-GRFs-FWI}
\end{table*}
\begin{figure}[!ht]
    \centering
    \includegraphics[width=0.5\linewidth]{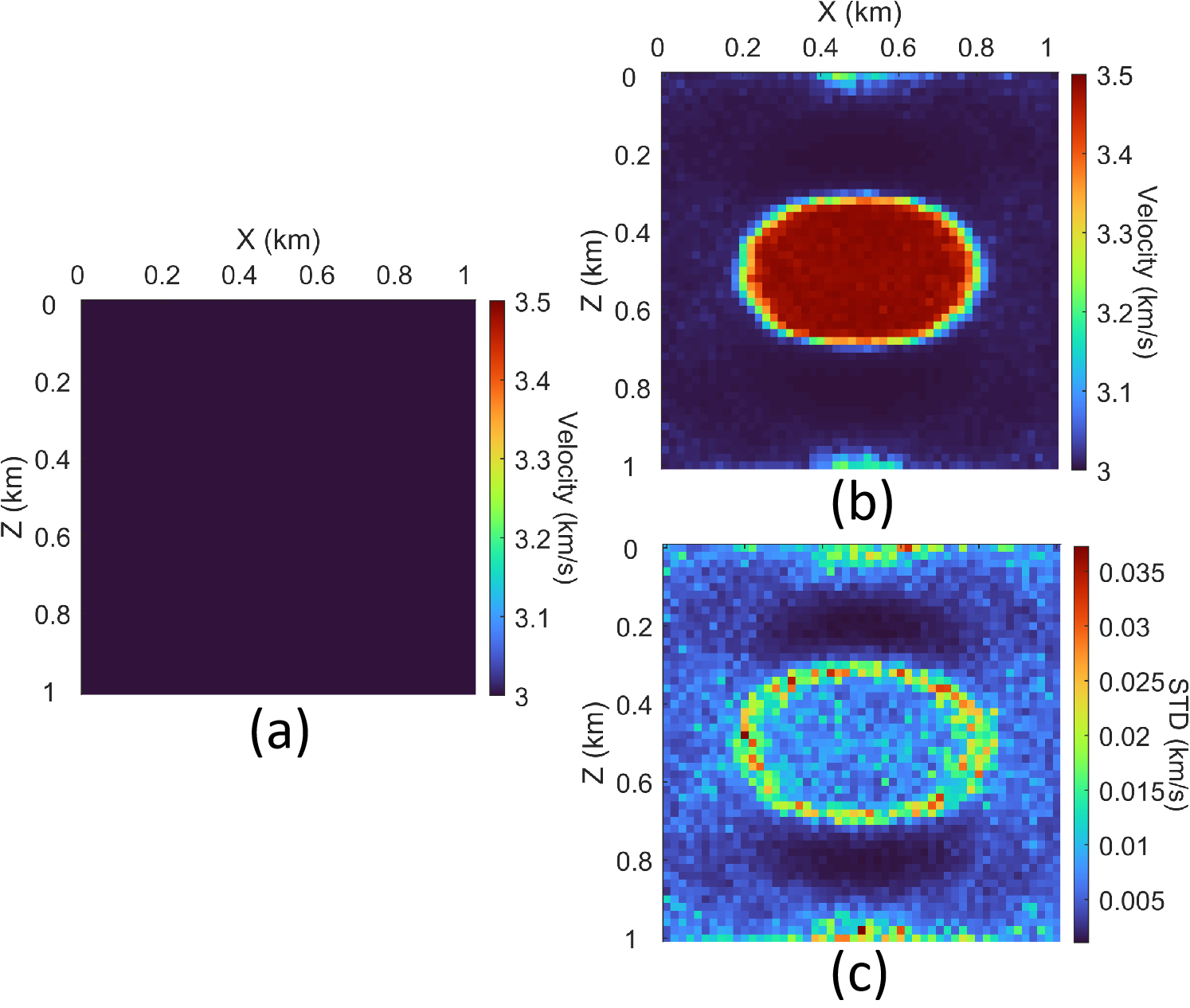}

    \caption{The inverse result of the inclusion model:  (a) the homogeneous initial model, (b)  the estimated velocity fields by the MCMC method using (a) as the initial model, and (c) the corresponding posterior standard deviations. }
    \label{fig:MCMC-Inclusion-result}
\end{figure}
\begin{figure}[!ht]
    \centering
    \includegraphics[width=0.7\linewidth]{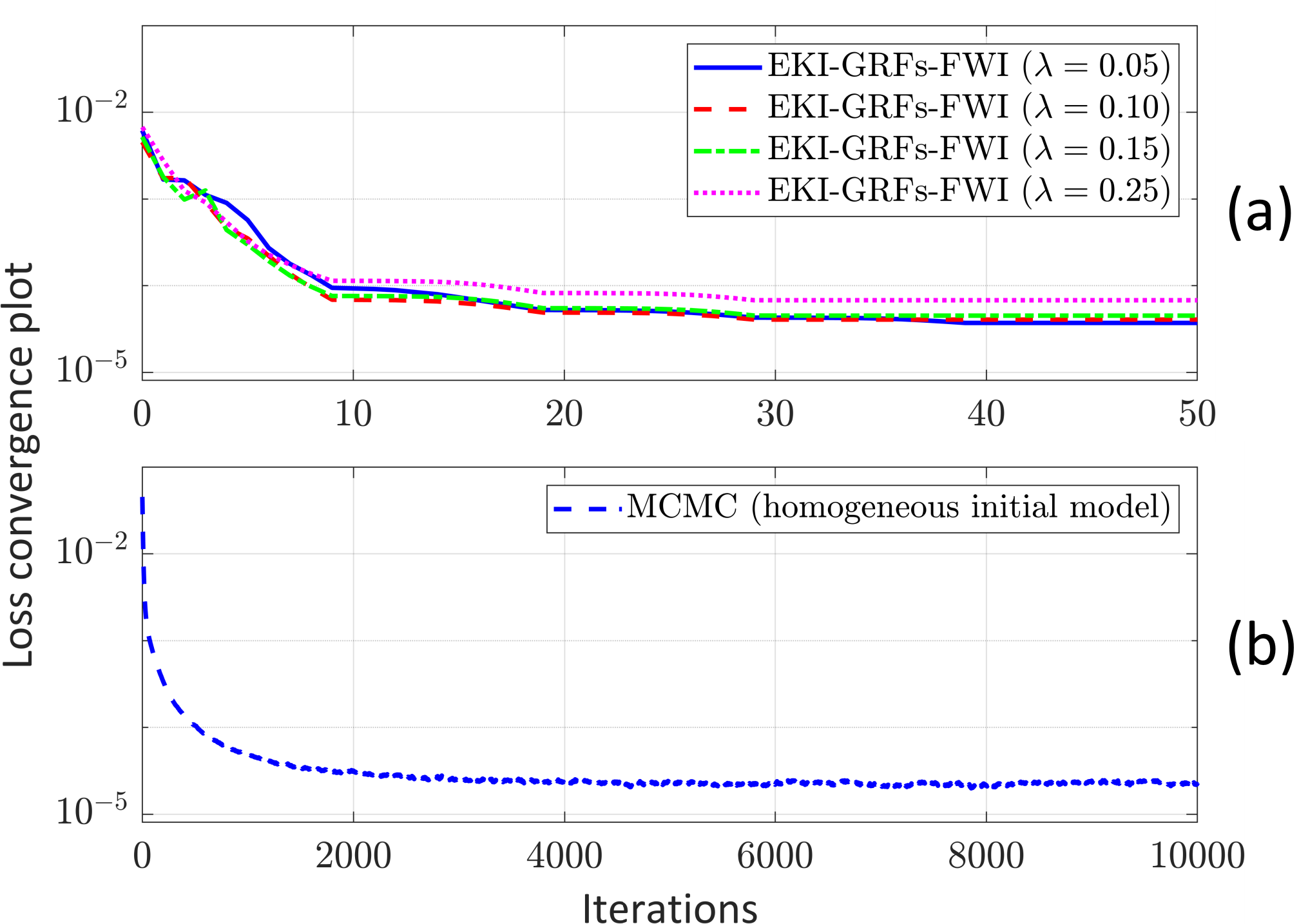}
   
    \caption{The loss convergence plot for inclusion model: (a) the EKI-GRFs-FWI method with different length-scale $\lambda$ and (b) the MCMC method with the homogeneous initial model.  }
    \label{fig:EKI-GRFs-FWI-Inclusion-convergence-plot}
\end{figure}
Fig.~\ref{fig:EKI-GRFs-FWI-Inclusion-result} (b), (d), (f) and (h) show the estimated velocity fields by EKI-GRFs-FWI for the inclusion model, 
where the length-scale $\lambda=0.05, 0.1,0.15$ and $0.25$ as in \eqref{eq:matern}, respectively. We observe that the elliptical shape can be recovered satisfactorily  
by EKI-GRFs-FWI when $\lambda=0.1$ and $0.15$ as shown in Fig.~\ref{fig:EKI-GRFs-FWI-Inclusion-result} (d) and (f). In contrast, Fig.~\ref{fig:EKI-GRFs-FWI-Inclusion-result} (b) and (h) show relatively poor velocity fields.  
The reason is that a relatively small length-scale ($\lambda = 0.05$) is used by EKI-GRFs-FWI, which results in a relatively rough estimated velocity field. When using a relatively large length-scale ($\lambda = 0.25$), EKI-GRFs-FWI leads to a relatively smooth estimated velocity field. 
The results demonstrate that EKI-GRFs-FWI can infer an accurate velocity field when adopting length-scale $\lambda=0.1$ and $0.15$ for the inclusion model, and the relative errors are about $1.56\%$ and $1.65\%$, respectively, as shown in Tab.~\ref{table:error-metric-and-time-EKI-GRFs-FWI}. 

Fig.~\ref{fig:EKI-GRFs-FWI-Inclusion-result} (c), (e), (g) and (i) show their corresponding posterior standard deviations. It is observed that the large standard deviation values mainly locate at 
the interface between the inclusions and the background. 
The results suggest that the larger uncertainty estimates correlate well with the larger estimated errors, which demonstrates the effectiveness of uncertainty estimates by EKI-GRFs-FWI.

Fig.~\ref{fig:MCMC-Inclusion-result} (b) presents the estimated velocity fields by MCMC method for the inclusion model, and the 
homogeneous velocity model shown in Fig.~\ref{fig:MCMC-Inclusion-result} (a) is served as the initial model.
In this example, the MCMC method is performed with 10,000 iterations, with a burn-in period of 5,000 step. The step size is set to $2\times10^{-4}$.
According to Fig.~\ref{fig:MCMC-Inclusion-result} (b), we observe that the MCMC method successfully recovers the elliptical shapes. The corresponding relative
error is about $1.46\%$ as in Tab.~\ref{table:error-metric-and-time-EKI-GRFs-FWI}.   Fig.~\ref{fig:MCMC-Inclusion-result} (c) shows the corresponding posterior standard deviations. It can be observed that larger standard deviation values are also concentrated around the interface between the elliptical shapes and the background, which coincides with the differences between the predicted velocity fields and the ground truth one. The results suggest that the EKI-GRFs-FWI method provides comparable uncertainty estimations to the MCMC method.

Fig.~\ref{fig:EKI-GRFs-FWI-Inclusion-convergence-plot} (a) and (b) show the loss convergence plots for EKI-GRFs-FWI and MCMC methods, which is calculated by \eqref{eq:stopping-criterion-1}. It can be observed that both methods are fully converged. According to Tab.~\ref{table:error-metric-and-time-EKI-GRFs-FWI}, MCMC method perform 10,000 iterations, resulting in a total computational time of 21 min and 16 s. In contrast, EKI-GRFs-FWI is stopped at the 40th, 32nd, 32nd, and 32nd iterations based on the stopping criterion in \eqref{eq:stopping-criterion-3} when using $\lambda = 0.05, 0.1, 0.15$, and $0.25$, respectively. The corresponding maximum computational time is only 54 s. Although the MCMC method is slightly more accurate, EKI-GRFs-FWI achieves comparable results in significantly less time.

\subsection{Checkerboard model}
\begin{figure}[!ht]
    \centering
    \includegraphics[width=0.4\linewidth]{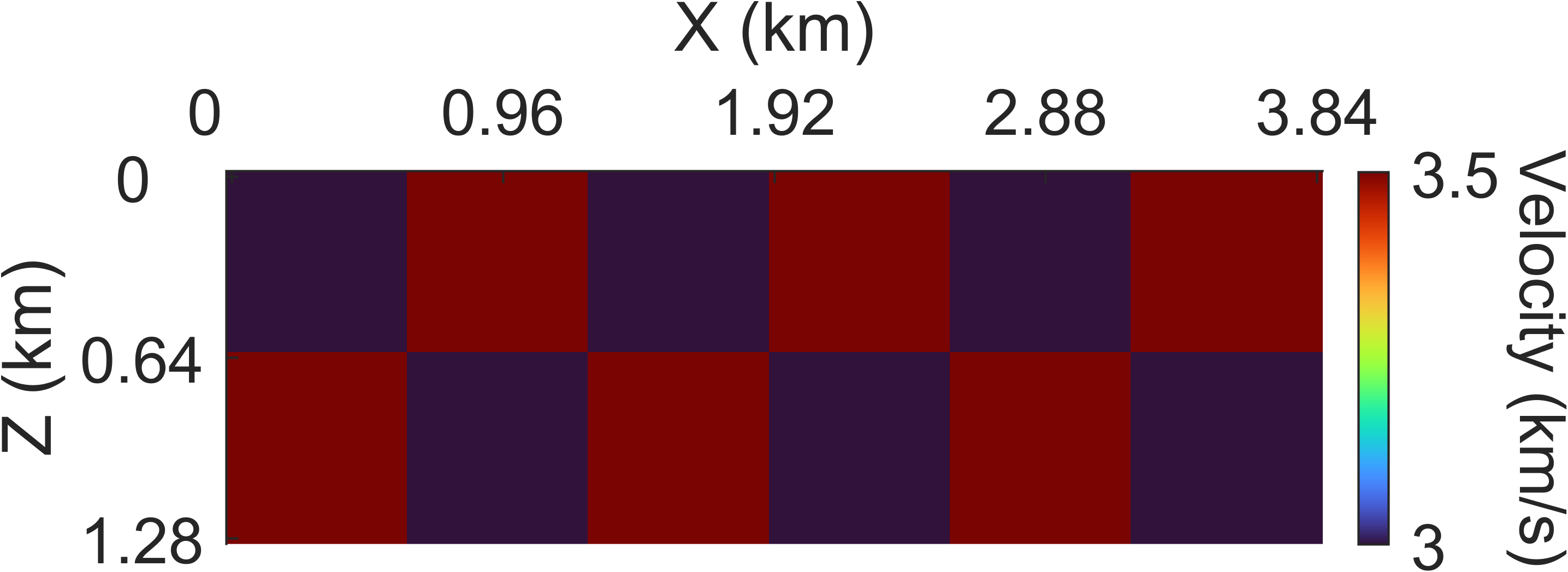}
    \caption{The checkerboard velocity model.  }
    \label{fig:EKI-GRFs-FWI-Checkerboard-model}
\end{figure}

Subsequently, we apply EKI-GRFs-FWI to the checkerboard velocity model as shown in Fig.~\ref{fig:EKI-GRFs-FWI-Checkerboard-model}. The size of velocity model is 3.84 km $\times$ 1.28 km with grid resolution of 40 m, which results in the mesh size of 97 $\times$ 33. For the acquisition system, 17 sources and 97 receivers are uniformly located at the surface. We infer the velocity fields using the observed data with frequencies $f^{(k)}=[2, 3, 4, 5, 6, 7, 8, 9, 10, 11, 12]$ Hz. For the EKI hyperparameters, we set the ensemble size $J=1000$ and stepsize $h=0.5$. 
\begin{figure*}[!ht]
    \centering
    \includegraphics[width=0.99\linewidth]{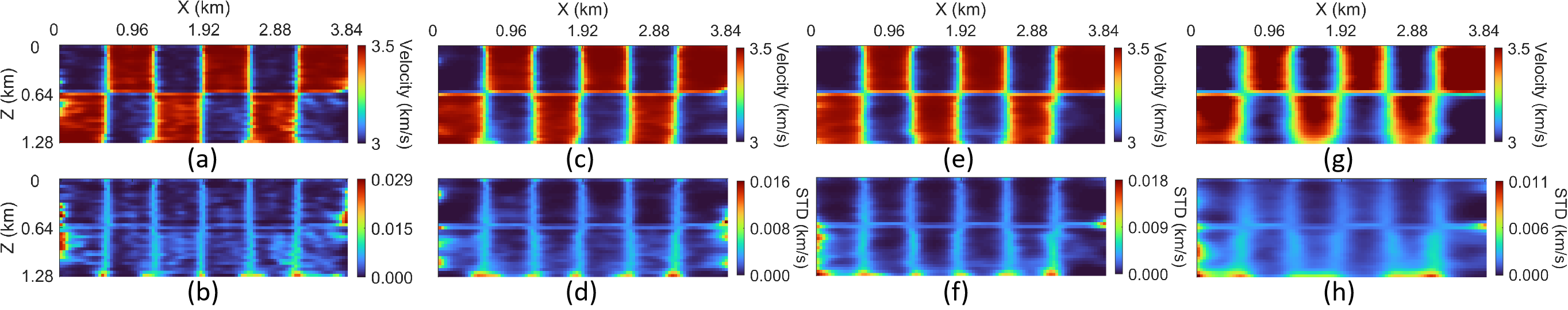}
    \caption{The result of EKI-GRFs-FWI with different length-scale $\lambda$ for checkerboard model: (a) and (b) the estimated velocity fields and its posterior standard deviations when $\lambda=0.05$; (c) and (d) the estimated velocity fields and its posterior standard deviations when $\lambda=0.1$; (e) and (f) the estimated velocity fields and its posterior standard deviations when $\lambda=0.15$; (g) and (h) the estimated velocity fields and its posterior standard deviations when $\lambda=0.25$.  }
    \label{fig:EKI-GRFs-FWI-Checkerboard-result}
\end{figure*}
\begin{figure}[!ht]
    \centering
    \includegraphics[width=0.6\linewidth]{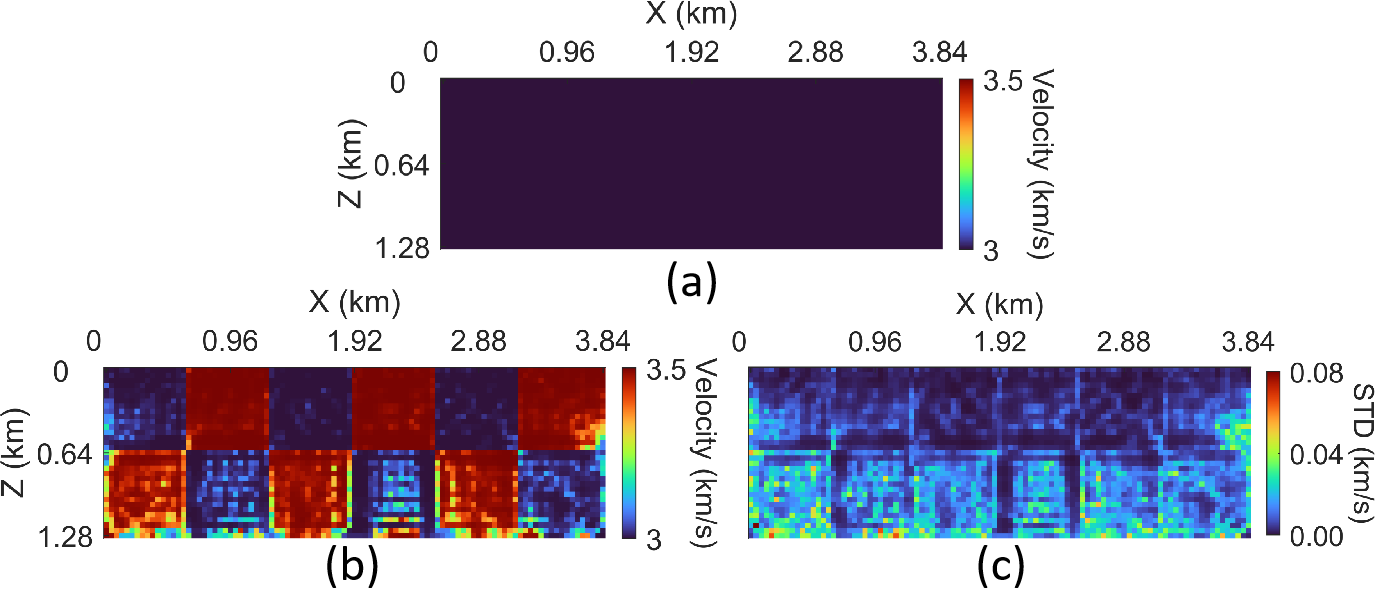}
   
    \caption{The inverse result of the checkerboard model:  (a) the homogeneous initial model, (b)  the estimated velocity fields by the MCMC method using (a) as the initial model, and (c) the corresponding posterior standard deviations. }
    \label{fig:MCMC-Checkerboard-result}
\end{figure}
\begin{figure}[!ht]
    \centering
    \includegraphics[width=0.7\linewidth]{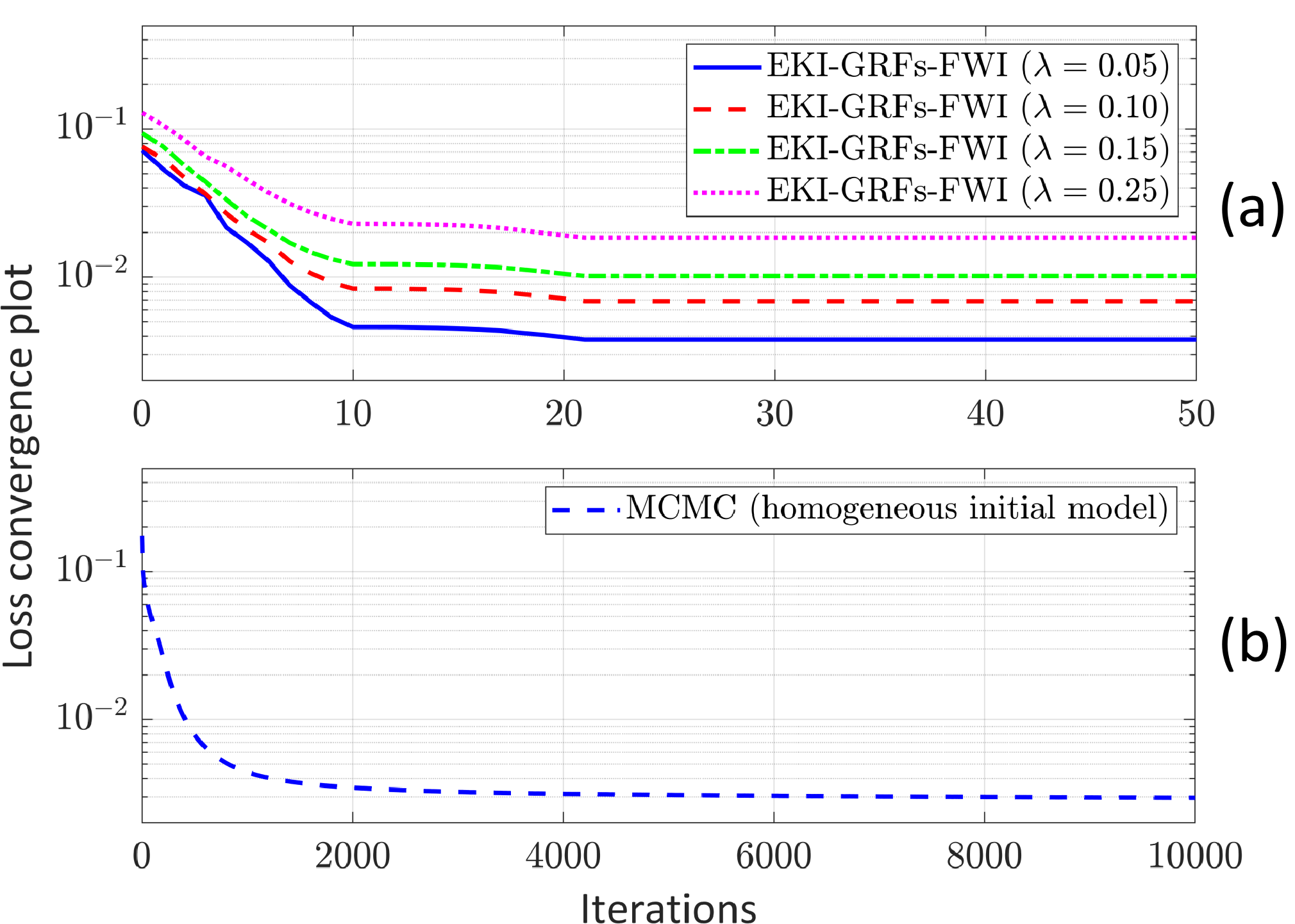}
   
    \caption{The loss convergence plot for checkerboard model: (a) the EKI-GRFs-FWI method with different length-scale $\lambda$ and (b) the MCMC method with the homogeneous initial model.  }
    \label{fig:EKI-GRFs-FWI-Checkerboard-convergence-plot}
\end{figure}
For this model, Fig.~\ref{fig:EKI-GRFs-FWI-Checkerboard-result} (a), (c), (e) and (g) show the estimated velocity fields by EKI-GRFs-FWI with $\lambda=0.05, 0.1,0.15$ and $0.25$, respectively. 
We observe that Fig.~\ref{fig:EKI-GRFs-FWI-Checkerboard-result} (a), (c) and (e)  approximately recover the shape of checkerboard.
In contrast, Fig.~\ref{fig:EKI-GRFs-FWI-Checkerboard-result} (g) presents a smooth velocity fields, especially in the deep region. 
Therefore, EKI-GRFs-FWI with $\lambda=$ 0.05, 0.10 and 0.15 can infer a relatively reasonable velocity field for the checkerboard model. According to Tab.~\ref{table:error-metric-and-time-EKI-GRFs-FWI}, the relative error of the estimated velocity fields by EKI-GRFs-FWI with $\lambda$= 0.05, 0.1 and 0.15 are about 2.15\%, 2.33\% and 2.50\%, respectively. 

From the posterior standard deviations shown in Fig.~\ref{fig:EKI-GRFs-FWI-Checkerboard-result} (b), (d), (f), and (h), we observe that large STD values are primarily distributed along the left, right and bottom boundaries of the model, as well as the interfaces in the model. 
The corresponding regions are not well recovered as shown on Fig.~\ref{fig:EKI-GRFs-FWI-Checkerboard-result} (a), (c), (e), and (g).
The results further demonstrate the effectiveness of the uncertainty estimations provided by EKI-GRFs-FWI.

Fig.~\ref{fig:MCMC-Checkerboard-result} (b) shows the estimated velocity fields by MCMC method for the checkerboard model, the corresponding initial model is shown in Fig.~\ref{fig:MCMC-Checkerboard-result} (a). 
The MCMC method is performed with 10,000 iterations, with a burn-in period of 5,000 step. The step size is set to $2\times10^{-4}$.
According to Fig.~\ref{fig:MCMC-Checkerboard-result} (b), we observe that MCMC method also recovers the shape of the checkerboard.
However, the deeper part ($Z=0.64$ km to $1.28$ km) of the model appears relatively rough. From Fig.~\ref{fig:MCMC-Checkerboard-result} (c), the regions with large STD values approximately corresponds to the rough spots as in Fig.~\ref{fig:MCMC-Checkerboard-result} (b). 
According to Tab.~\ref{table:error-metric-and-time-EKI-GRFs-FWI}, the relative error of estimated velocity fields by the MCMC method is about 2.53\%. 

Fig.~\ref{fig:EKI-GRFs-FWI-Checkerboard-convergence-plot} (a) and (b) exhibit the loss convergence plots for EKI-GRFs-FWI and MCMC methods, indicating that they are fully converged.
The MCMC method takes approximately 31 min and 40 s during the inversion procedure. In contrast, the maximum computational time of EKI-GRFs-FWI is only 1 min and 41 s. The results demonstrate that EKI-GRFs-FWI is significantly computationally efficient. 

\subsection{Overthrust model}
\begin{figure}[!ht]
    \centering
    \includegraphics[width=0.6\linewidth]{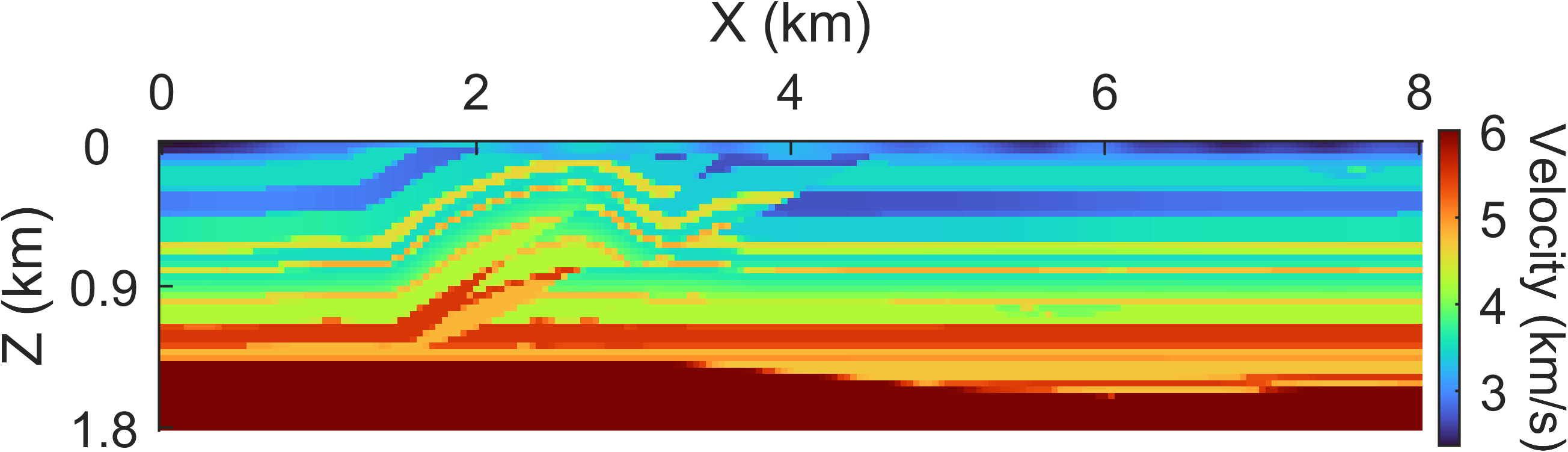}
    \caption{The overthrust velocity model.  }
    \label{fig:EKI-GRFs-FWI-Overthrust-model}
\end{figure}

Finally, we validate the performance of EKI-GRFs-FWI for the overthrust model as shown in Fig.~\ref{fig:EKI-GRFs-FWI-Overthrust-model}.  The size of the velocity model is 8~km $\times$ 1.8~km and the grid size is 40 m. Consequently, its mesh size is 201 $\times$ 46.
In this case, 21 sources and 201 receivers are evenly placed at the surface. The velocity model is inferred using the observed data with frequencies $f^{(k)}=[2, 2.5, 3, 4, 5, 7, 9, 11,13,15]$ Hz. For EKI hyperparameters, the ensemble size $J$ is set to 3000 with a stepsize of $h=0.5$. 
\begin{figure*}[!ht]
    \centering
    \includegraphics[width=0.99\linewidth]{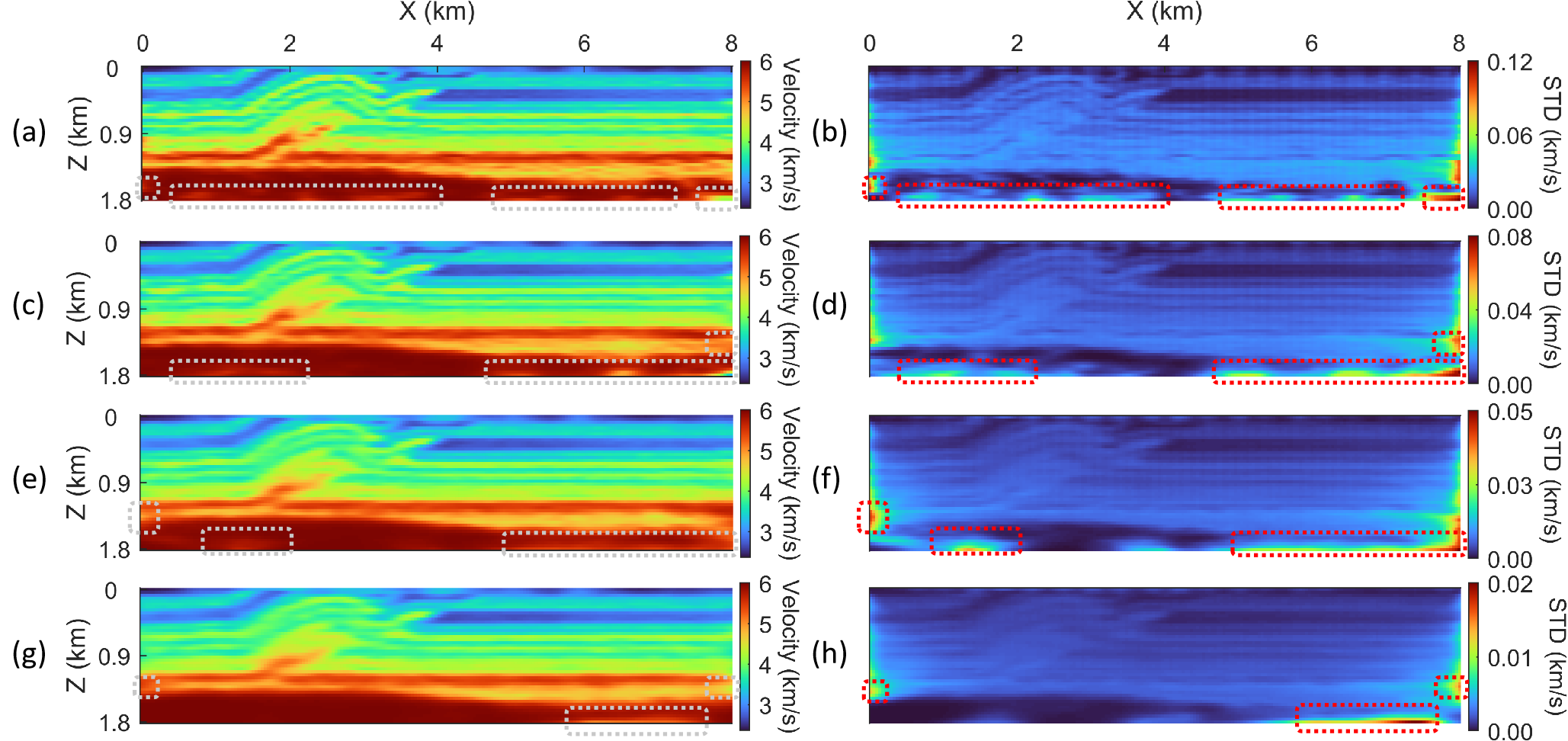}
    \caption{The result of EKI-GRFs-FWI with different length-scale $\lambda$ for overthrust model: (a) and (b) the inferred velocity fields and its posterior standard deviations when $\lambda=0.05$; (c) and (d) the inferred velocity fields and its posterior standard deviations when $\lambda=0.1$; (e) and (f) the inferred velocity fields and its posterior standard deviations when $\lambda=0.15$; (g) and (h) the inferred velocity fields and its posterior standard deviations when $\lambda=0.25$.  }
    \label{fig:EKI-GRFs-FWI-Overthrust-result}
\end{figure*}
\begin{figure*}[!ht]
    \centering
    \includegraphics[width=0.99\linewidth]{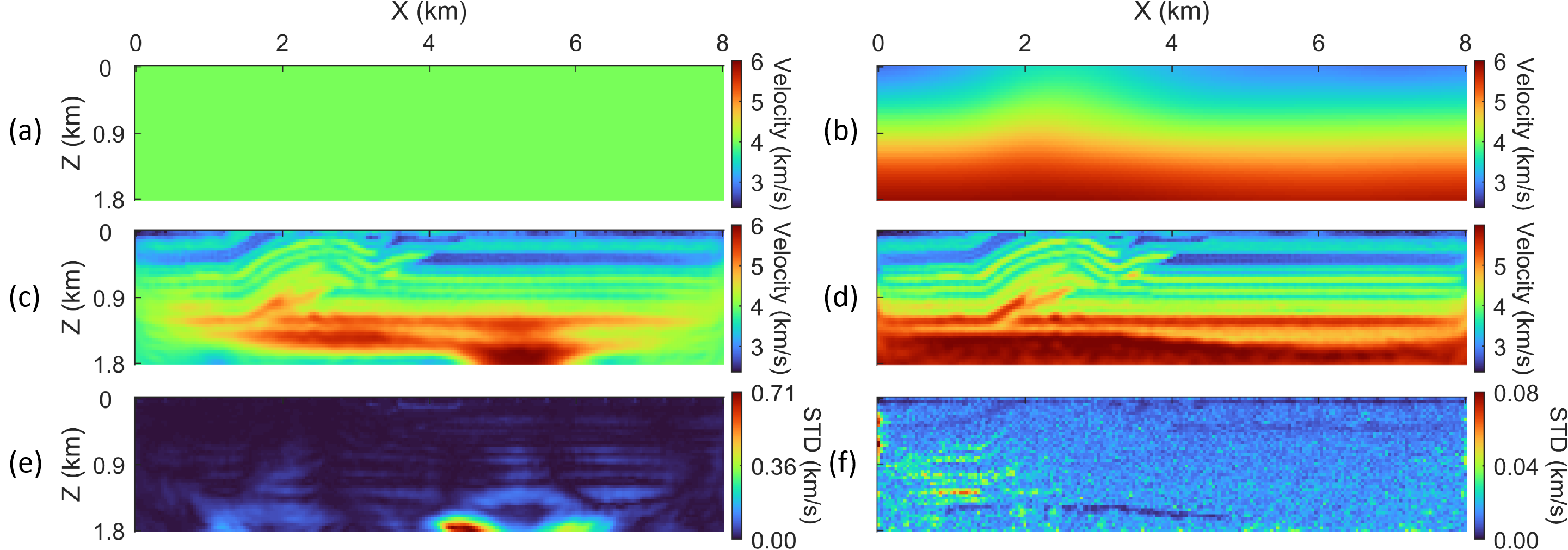}

    \caption{The inverse result of the overthrust model: (a) the homogeneous initial model, (c) and (e) the estimated velocity fields and the corresponding posterior standard deviations by the MCMC method using (a) as the initial model, (b) the smoothed initial model, and (d) and (f) the estimated velocity fields and the corresponding posterior standard deviations by the MCMC method using (b) as the initial model.  }
    \label{fig:MCMC-Overthrust-result}
\end{figure*}
\begin{figure}[!ht]
    \centering
    \includegraphics[width=0.7\linewidth]{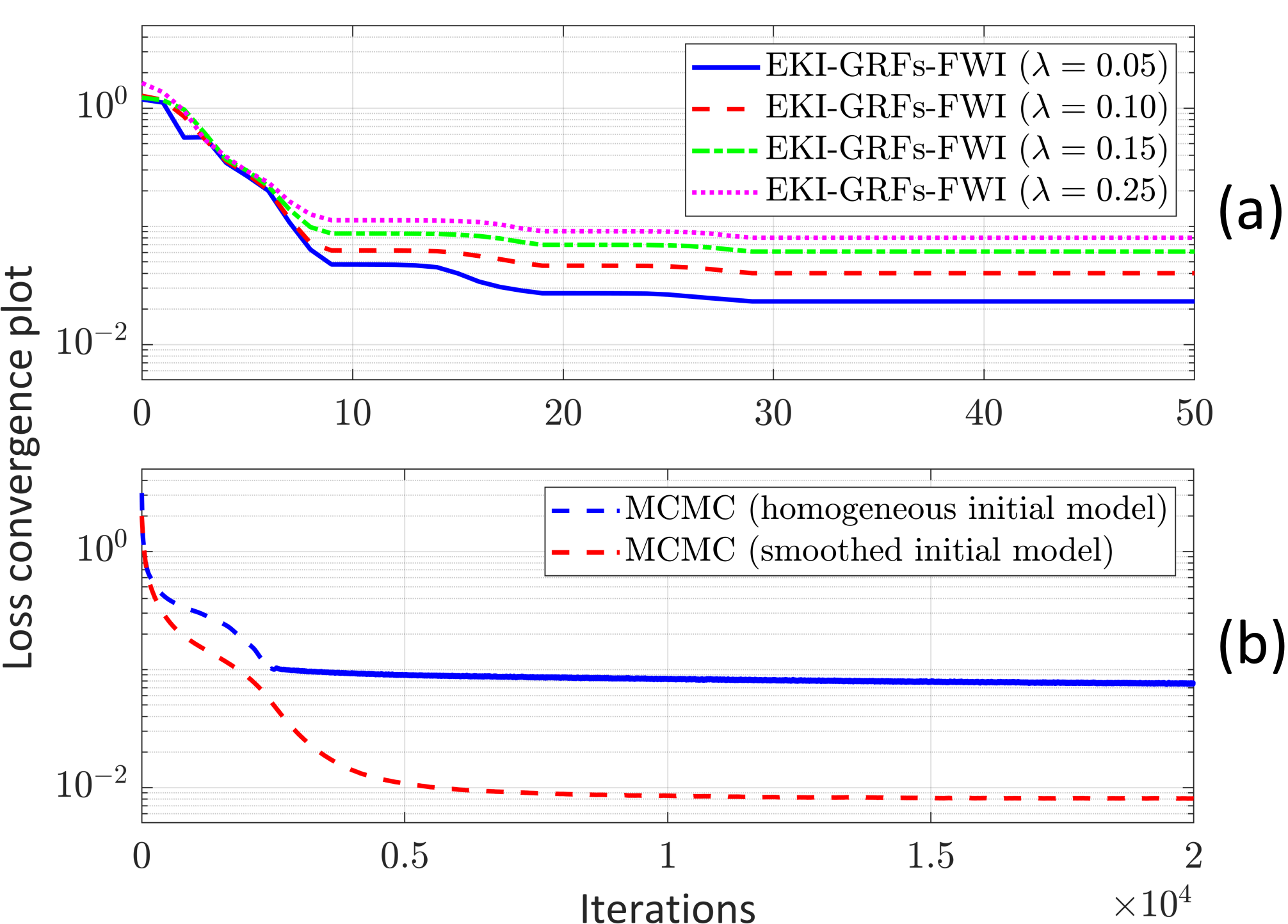}
   
    \caption{The loss convergence plot for overthrust model: (a) the EKI-GRFs-FWI method with different length-scale $\lambda$ and (b) the MCMC method with different initial model.  }
    \label{fig:EKI-GRFs-FWI-Overthrust-convergence-plot}
\end{figure}
Fig.~\ref{fig:EKI-GRFs-FWI-Overthrust-result} (a), (c), (e) and (g) show the estimated velocity fields for the overthrust model, which is obtained by EKI-GRFs-FWI with $\lambda=0.05, 0.1,0.15$ and $0.25$, respectively. Despite the complexity of this model, EKI-GRFs-FWI with different length-scales can approximately capture its complex features in the shallow ($Z=$ 0~km to 0.9~km) layers. However, as the length-scale increases, the interfaces between these features gradually become more blurred. 
The result suggests that for complex model such as the overthrust model, adopting a small length-scale value (e.g., $\lambda=0.05$) can help  EKI-GRFs-FWI method to better capture the detailed features of the velocity model.

Fig.~\ref{fig:EKI-GRFs-FWI-Overthrust-result} (b), (d), (f) and (h) show the corresponding posterior standard deviations. The large STD values are mainly distributed along the left, right and bottom boundaries of the model. Correspondingly, Fig.~\ref{fig:EKI-GRFs-FWI-Overthrust-result} (a), (c), (e), and (g) show that the regions with large STD values may not be sufficiently recovered (see the gray and red dashed boxes in Fig.~\ref{fig:EKI-GRFs-FWI-Overthrust-result}). 

Fig.~\ref{fig:MCMC-Overthrust-result} (c) and (d) illustrate the estimated velocity fields by MCMC method for the overthrust model, where Fig.~\ref{fig:MCMC-Overthrust-result} (a) and (b) are served as the initial models, respectively. For overthrust model, MCMC method is performed with 20,000 iterations, including a burn-in period of 10,000 steps, and the step size is $2\times10^{-4}$. Fig.~\ref{fig:MCMC-Overthrust-result} (d) shows more accurate results compared to Fig.~\ref{fig:MCMC-Overthrust-result} (c). The reason is that Fig.~\ref{fig:MCMC-Overthrust-result} (b) shows a smooth version of the overthrust model, which includes more low frequency information than the homogeneous initial model as shown in Fig.~\ref{fig:MCMC-Overthrust-result} (a). 
This result suggests that the MCMC method strongly relies on the initial velocity model. 
Fig.\ref{fig:MCMC-Overthrust-result} (f) shows the posterior standard deviations by MCMC method using Fig.~\ref{fig:MCMC-Overthrust-result} (b) as the initial model. We observe that the large standard deviation values are primarily located at the interfaces in the left areas ($X=$ 0~km to 2~km) of the model, suggesting that the estimated velocities at these interfaces may be inaccurate.

Fig.~\ref{fig:EKI-GRFs-FWI-Overthrust-convergence-plot} show the loss convergence plots for EKI-GRFs-FWI and MCMC methods, which indicates that they are converged.
According to Tab.~\ref{table:error-metric-and-time-EKI-GRFs-FWI}, when using Fig.~\ref{fig:MCMC-Overthrust-result} (b) as the initial model, the relative error of the estimated velocity fields by MCMC method is about 4.91\%. The relative error of the estimated velocity fields by the EKI-GRFs-FWI with $\lambda=0.05$ is about 4.93\%.  The results suggest that the EKI-GRFs-FWI method provides comparable estimated velocity fields as those of the MCMC method. Additionally, the maximum computational time of EKI-GRFs-FWI is only 18 min and 40 s, which is significantly efficient than that of MCMC method (approximately 2 h and 25 min).  It demonstrates that EKI-GRFs-FWI method can achieve a balance between computational efficiency and accuracy of the estimated velocity fields.
\section{Conclusion}
This paper presents an efficient uncertainty-aware frequency-domain acoustic full waveform inversion (FWI) method that integrates Gaussian Random Fields (GRFs) with the ensemble Kalman inversion (EKI) algorithm. The proposed approach simultaneously infers subsurface velocity fields and quantifies the associated uncertainty in the inversion results.

A key strength of the method lies in leveraging the derivative-free nature and rapid convergence of EKI, which significantly contributes to its computational efficiency. By incorporating prior knowledge of spatial smoothness and correlation length scales, GRFs are used to generate high-quality initial ensembles for EKI, thereby improving the accuracy of the inferred velocity models. To assess
the performance of the proposed method, we compare EKI-GRFs-FWI method with a gradient-based MCMC method on three benchmark problems. Experimental results demonstrate that the EKI-GRFs-FWI method is more computationally efficient compared to MCMC method. Furthermore, EKI-GRFs-FWI can strike a balance between computational efficiency and accuracy of the inferred velocity fields, while also providing informative uncertainty estimate about the inferred results.

\bibliographystyle{ieeetr}   
\bibliography{references}

\end{document}